\RequirePackage{amsmath}
\documentclass[10pt]{iopart}
\usepackage{iopams}
\usepackage[dvips]{graphicx}
\usepackage{cite}

\DeclareMathOperator{\sgn}{sgn}
\DeclareMathOperator{\Hev}{H}
\DeclareMathOperator{\ceil}{ceil}

\begin{document}

\title[]{Analytical derivation of DC SQUID response}

\author{I I Soloviev$^{1,2,3}$, N V Klenov$^{1,3,4,5}$, A E Schegolev$^{4}$, S V Bakurskiy$^{1,3}$ and M Yu Kupriyanov$^{1,3,6}$}

\address{$^{1}$Lomonosov Moscow State University Skobeltsyn Institute of Nuclear Physics, 119991, Moscow, Russia}
\address{$^{2}$Lukin Scientific Research Institute of Physical Problems, Zelenograd, 124460, Moscow, Russia}
\address{$^{3}$Moscow Institute of Physics and Technology, State University, Dolgoprudniy, Moscow region, Russia}
\address{$^{4}$Physics Department, Moscow State University, 119991, Moscow, Russia}
\address{$^{5}$N. L. Dukhov All-Russia Research Institute of Automatics, 127055, Moscow, Russia}
\address{$^{6}$Solid State Physics Department, Kazan Federal University, 420008, Kazan, Russia}

\ead{isol@phys.msu.ru} \vspace{10pt}

\begin{abstract}
We consider voltage and current responses formation in DC SQUID with
overdamped Josephson junctions in resistive and superconducting
state in the frame of resistively shunted junction (RSJ) model. For
simplicity we neglect the junction capacitance and the noise effect.
Explicit expressions for the responses in resistive state were
obtained for a SQUID which is symmetrical with respect to bias
current injection point. Normalized SQUID inductance $l = 2 e I_c
L/\hbar$ (where $I_c$ is the critical current of Josephson junction,
$L$ is the SQUID inductance, $e$ is the electron charge and $\hbar$
is the Planck constant) was assumed to be within the range $l \leq
1$, subsequently expanded up to $l \approx 7$ using two fitting
parameters. SQUID current response in superconducting state was
considered for arbitrary value of the inductance. Impact of small
technological spread of parameters relevant for low-temperature
superconductor (LTS) technology was studied with generalization of
the developed analytical approach for a case of small difference of
critical currents and shunt resistances of the Josephson junctions,
and inequality of SQUID inductive shoulders for both resistive and
superconducting states. Comparison with numerical calculation
results shows that developed analytical expressions can be used in
practical LTS SQUIDs and SQUID-based circuits design, e.g. large
serial SQIF, drastically decreasing the time of simulation.
\end{abstract}

\pacs{85.25.Dq, 85.25.Am}

\vspace{2pc} \noindent{\it Keywords}: DC SQUID, voltage response,
current response, SQIF
%
%
\maketitle
%
\ioptwocol

\section{Introduction}
The superconducting quantum interference device (SQUID) is a basic
component of superconductor electronics having numerous applications
\cite{CB, W, GV}. DC SQUID being basically a magnetic
flux-to-voltage transformer is used e.g. in highly sensitive
magnetometers \cite{SZSSALMM, CSSZMM, SSZSMAFM, SSZSAFMMM, SSZAFRM},
amplifiers \cite{ProM, SSZSAFM, KSKMSQIFDriv}, readout circuits
\cite{ZSKHAPMKSESM} and antennas \cite{KSSKM,AETBH}. These devices
are routinely designed for fabrication process utilizing
low-temperature superconductors (LTS) and tunnel Josephson
junctions, which became a workhorse of modern superconducting
electronics.

High accuracy of modern LTS fabrication technology allow for the
construction of advanced SQUID-based structures with unconventional
flux-to-voltage transformation. One example of such structures is
Superconductor Quantum Array (SQA) \cite{KSSKSM} with highly linear
voltage response. SQA can be based on superconducting quantum
interference filters (SQIFs) resulting from SQUID arrays with
unequal loops \cite{OHS, HOS}, or bi-SQUID cells \cite{KSKM09}. It
was argued \cite{KSKM11} that SQA can be a basis of active
electrically small wideband superconductor antenna. Such antenna
could provide significant advances to modern superconducting
broadband radio-frequency receiving systems \cite{MKVFK}. A variety
of structures, like the ones mentioned above, were proposed and
extensively studied both theoretically and experimentally in recent
years
\cite{KSSKM,AETBH,KSKM10,KSKM11,KSKM09,KSSKSM,KSKSM,LBEPR11,SSKSS,LBEPR,PMETA,BPLPM,WCAD,KSSM}.

Qualitative understanding of SQUID or SQUID-based structure response
\cite{OHS, HOS, KSKM09, KSSM} can be based on analytical approach
assuming the SQUID inductance to be approximately equal to zero
\cite{BP, L}. However, for quantitative estimation of designed LTS
circuit characteristics, accounting for a certain real value of the
inductance is inevitable.

Main effort in the development of time-averaged analytical
dependencies for practical SQUID parameters was done during study of
high-temperature superconductor (HTS) SQUIDs
\cite{C98,C99,C991,G02,G03,GNSM}. High noise level $\Gamma > 0.1$
($\Gamma = 2\pi k_B T/I_c \Phi_0$, where $k_B$ is the Boltzmann
constant, $I_c$ is the critical current of Josephson junction,
$\Phi_0 = h/2e$ is the flux quantum, $h$ is the Planck constant, $e$
is the electron charge) and high value of SQUID inductance $L
> 100$~pH are typical in this case, that results in $l\Gamma \geq 1$
($l = 2\pi L I_c/\Phi_0$). Developed analytical approaches are valid
for $l\Gamma \geq 1$ and based on solution of two-dimensional
Fokker-Planck equation \cite{C98,G02}. Obtained expression for the
voltage-flux function $\overline{v}(\phi_e)$ of symmetrical SQUID
(where $\overline{v}$ is the average voltage across the SQUID, and
$\phi_e = \pi\Phi_e/\Phi_0$, $\Phi_e$ is the external magnetic flux)
in the frame of these approaches is a simple harmonic function
$\overline{v} = a\cos\phi_e + b$ (where $a,b$ are constants), which
is not intended for accurate description of the response shape.
These methods are used mostly for estimation of the voltage response
amplitude and corresponding maximum value of the transfer function.
Analysis of the presented results done in Ref.~\cite{G02} shows that
the approaches can't be applied for LTS SQUIDs.

Perturbation analysis developed in the limit of small SQUID
inductance $l$ was shown to be well suited for study of some aspects
of SQUID dynamics and characteristics like thermal escape problem
\cite{GJTC}, Shapiro steps \cite{RD}, persistent current and
magnetic susceptibility \cite{TDL}. Though, study of time-averaged
response for practical inductance values still requires numerical
calculations \cite{DLFG}.

It is interesting to note that an attempt to average SQUID voltage
dynamics in the frame of another analytical approach, which is
somewhat similar to the perturbation analysis, was undertaken even
before the works \cite{C98,G02} on HTS SQUIDs. In 1983 Peterson and
McDonald proposed their voltage and current expressions for DC SQUID
with small but finite inductance \cite{PM}. In their method the
SQUID difference phase $\psi = (\phi_1 - \phi_2)/2$ ($\phi_{1,2}$
are the Josephson phases of the first and the second junction) was
expressed as a sum of the normalized external magnetic flux $\phi_e$
and a small correction $x$, which vanishes with the inductance,
$\psi = - \phi_e + x$. The correction $x$ was sought in a form of
Fourier series, and so all final analytical expressions also contain
a sum of certain series. This complexity complicates analysis of
obtained solutions and their adaptation for more complex practical
SQUID-based circuits like a bi-SQUID or a SQIF.

Despite of the fact that half a century is gone since the year of
the first demonstration of quantum interference between two
Josephson junctions connected in parallel by superconducting
inductance \cite{JLSM}, a shape of DC SQUID response still was not
found analytically for practical parameters of the device at low
temperature ($T \approx 4.2$~K). This statement is confirmed by the
fact that practical circuit optimization is always performed in the
frame of numerical analysis
\cite{KSKM11,KSKSM,LBEPR11,LBEPR,WCAD,KSSM} that slows down the
design process.

Purpose of this paper is resolution of this long standing problem
with presentation of analytical analysis of DC SQUID voltage and
current responses on applied magnetic flux in resistive state (see
Section 2 and Section 3 correspondingly), and analysis of the
current response in superconducting state (Section 3). The analysis
includes consideration of effect of small technological spread of
SQUID parameters relevant for LTS technology, which is presented in
Section 4. While superconducting state is considered for arbitrary
value of the inductance, resistive state is considered first for $l
\leq 1$. Analytical voltage-flux and current-flux functions in
resistive state for practical inductances up to $l \approx 7$ are
obtained by fitting of numerical data in Section 5. We conclude the
paper by discussion of applicability of the obtained analytical
expressions to optimization of practical SQUID-based structures like
SQIF. For simplicity we don't account for thermal noise and neglect
capacitance of the Josephson junctions. Their effects on the
responses of LTS DC SQUID with overdamped junctions having
relatively high critical currents ($\Gamma \leq 10^{-3}$) are
assumed to be small.

\section{SQUID voltage}

\begin{figure}[t]
\begin{center}
\includegraphics[width=4cm]{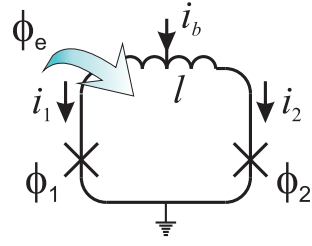}
\end{center}
\caption{Symmetrical DC SQUID fed by bias current $i_b$ and applied
magnetic flux $\phi_e$.} \label{Fig2}
\end{figure}

In this Section we consider the voltage response of symmetrical DC
SQUID with inductance $l \leq 1$ (Fig.~\ref{Fig2}). In the frame of
resistively shunted junction (RSJ) model \cite{RSJ} for overdamped
Josephson junctions one can write simple equations for the currents
$i_{1,2}$ flowing through the SQUID
\begin{align*}
i_{1,2} = \sin\phi_{1,2} + \dot{\phi}_{1,2}\nonumber,
\end{align*}
where the currents are normalized to the critical current $I_c$ and
dot denotes time differentiation with normalized time $\tau =
t\omega_c$, $\omega_c = 2\pi I_c R_n/\Phi_0$ is the characteristic
frequency, and $R_n$ is the shunt resistance of the junctions.
Kirchoff equations for the SQUID produce the following system of
differential equations:
\begin{subequations} \label{SQUIDeq}
\begin{equation}
\frac{l}{2}\dot{\psi} = -(\psi + \phi_e) -
\frac{l}{2}\sin\psi\cos\theta,
\end{equation}
\begin{equation}
\dot{\theta} = \frac{i_b}{2} - \cos\psi\sin\theta,
\end{equation}
\end{subequations}
where $\theta = (\phi_1 + \phi_2)/2$ is the sum phase, $i_b =
I_b/I_c$, $I_b \geq 0 $ is the bias current.

Since the inductance is present only in equation (\ref{SQUIDeq}a),
we assume that it primarily affects the difference phase. Following
works \cite{KSKM09,PM} we consider the difference phase as a sum of
slow-varying $\psi_-$ and oscillating $\psi_\sim$ parts. The last
one is assumed to be small $\psi_\sim \ll 1$ for shunted junctions
and the inductance value $l \leq 1$. SQUID voltage response
$\overline{v} = w_J$ normalized to $I_cR_n$ product (where $w_J$ is
the Josephson frequency) is found in the following sequence. First,
we obtain the solution $w_{J-} = \overline{\dot{\theta}}$ assuming
that $\psi = \psi_-$. Then we find $\psi_\sim$ from (\ref{SQUIDeq}a)
using the determined $\theta$, that in turn allows us to find the
correction $w_{J\sim}$ and the total Josephson frequency $w_J =
w_{J-} + w_{J\sim}$.

\subsection{Solution for $\psi = \psi_-$}

The difference phase is equal to it's slow-varying part in the limit
of vanishing inductance $l \rightarrow 0$. Equation (\ref{SQUIDeq}a)
in this case gives $\psi = \psi_- = -\phi_e$, and an according
solution of equation (\ref{SQUIDeq}b) is well known (see
Ref.~\cite{PM} and references therein):
\begin{equation} \label{SolFor0ap}
\tan\frac{\theta}{2} = z + \sqrt{1
-z^2}\tan\frac{\frac{i_b}{2}\sqrt{1-z^2}\tau}{2},
\end{equation}
where $z = (2/i_b)\cos\phi_e$. The Josephson oscillation frequency
can be obtained from (\ref{SolFor0ap}) directly
\begin{equation} \label{SQwj0}
w_{J-} = \frac{i_b}{2}\sqrt{1-z^2} = \sqrt{\frac{i_b^2}{4} -
\cos^2\phi_e}.
\end{equation}

It is worth to mention the known consistency of the expression
(\ref{SQwj0}) with the one for a single junction $w_J = \sqrt{i_b^2
- i_c^2}$, so one can treat $\cos\phi_e$ as an effective critical
current and $i_b/2$ as an effective bias current. $w_{J-}$ should be
put zero if $i_b/2 < |\cos\phi_e|$.

\subsection{Solution for $\psi = \psi_- + \psi_\sim$}

Substitution of the difference phase as the sum $\psi_- + \psi_\sim$
into (\ref{SQUIDeq}a) in the limit $\psi_\sim \ll 1$ converts this
equation into
\begin{equation} \label{SQx1Eq}
\frac{l}{2}\dot{\psi}_\sim = -\psi_\sim - \frac{l}{2}(\sin\psi_- +
\psi_\sim \cos\psi_-)\cos\theta.
\end{equation}
Solution of equation (\ref{SQx1Eq}) has a form
\begin{equation} \label{psivar}
\psi_\sim = \frac{l\sin\psi_-[l(\cos\psi_-
-\frac{i_b}{2}\sin\theta)- 2\cos\theta]}{l^2 w^2_{J-}+4}.
\end{equation}
Its substitution into (\ref{SQUIDeq}b) leads to correction of the
Josephson frequency
\begin{equation}
w_J = \frac{i_b}{2} - \cos\psi_-\overline{\sin\theta} +
\sin\psi_-\overline{\psi_\sim\sin\theta}.
\end{equation}
Explicit form for $w_J$ can be found by time averaging
\begin{equation}\label{SQwjtot1}
w_J = w_{J-} - \frac{l^2 w^2_{J-}}{l^2 w^2_{J-}+4}
\left(\frac{i_b}{2} - w_{J-}\right)\tan^2\psi_-.
\end{equation}
This expression describes SQUID voltage response for $l \leq 1$.

First factor of the second term in (\ref{SQwjtot1}) suggests that
decrease of the voltage response comes from filtering properties of
the SQUID. This can be understood qualitatively on example of
synchronization of the junctions by circulating current at external
flux equal to half flux quantum $\phi_e = \pi/2$.

In general, the circulating current is defined as $i_{cir} = (i_1 -
i_2)/2$. According to (\ref{SQUIDeq}a) in resistive state it is
$i_{cir} = -2\psi_\sim/l$ and with expression (\ref{psivar}) it
results to
\begin{multline}\label{icirvar}
i_{cir} = \frac{4\sqrt{1+\frac{l^2i_b^2}{16}}}{l^2w^2_{J-} +
4}\sin\psi_-\sin\left(\theta +
\arctan\left[\frac{4}{li_b}\right]\right)\\
-\frac{l\sin2\psi_-}{l^2w^2_{J-} + 4}.
\end{multline}

\begin{figure}[t]
\centering
\begin{minipage}[t]{0.49\columnwidth}
\resizebox{1\columnwidth}{!}{
\includegraphics{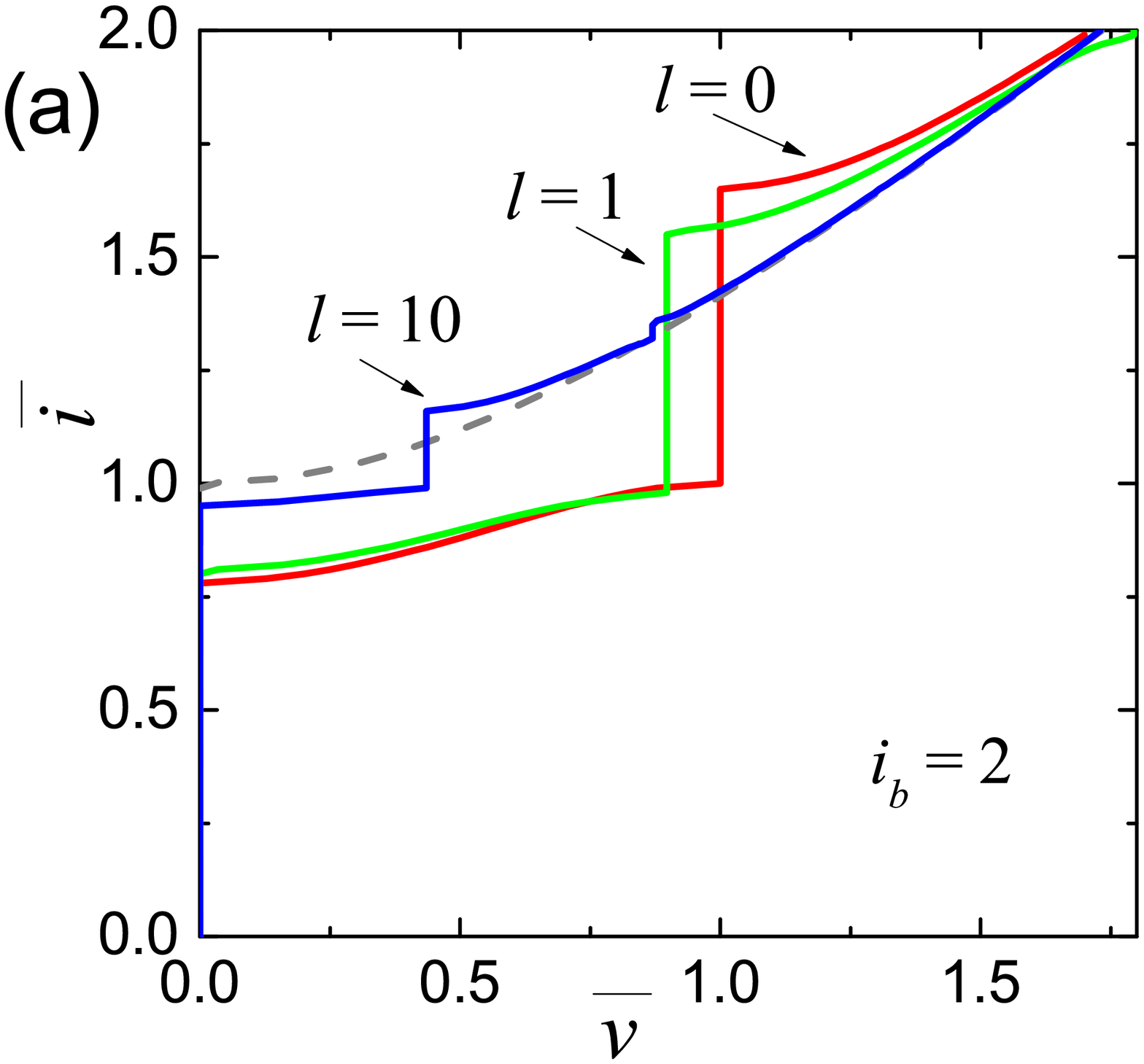}}
\end{minipage}
\begin{minipage}[t]{0.49\columnwidth}
\resizebox{1\columnwidth}{!}{
\includegraphics{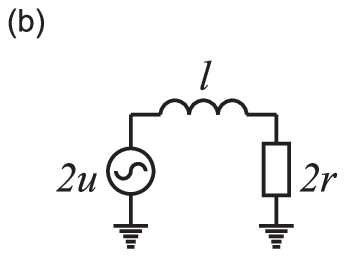}}
\end{minipage}
\caption{(a) IV curves of Josephson junction biased by current $i_b
= \bar{i} + i_{cir}^a\cos(w_J\tau)$, where $i_{cir}^a$ corresponds
to the amplitude of the circulating current and $w_J$ corresponds to
the Josephson frequency of a junction in the SQUID biased at $\phi_e
= \pi/2$. The parameters $i_{cir}^a$, $w_J$ are calculated for
$i_b/2 = 1$ by expressions (\ref{icirvar_J_a}), (\ref{maxV}) for $l
= 0,~1$, and calculated numerically using (\ref{SQUIDeq}) for $l =
10$. Dashed line corresponds to $i_{cir}^a = 0$. The IV curves are
calculated numerically using (\ref{SQUIDeq}). (b) $LR$-filter
circuit used to analyze the SQUID at $\phi_e = \pi/2$.} \label{Fig1}
\end{figure}

For the considered applied flux $\psi_- = -\phi_e = -\pi/2$ the sum
phase is $\theta = w_{J-}\tau$ and the frequency $w_{J-} = i_b/2$ as
it follows from (\ref{SolFor0ap}), (\ref{SQwj0}). Assuming that $l
\ll 1$ the circulating current can be presented as $i_{cir} \approx
-\cos(i_b\tau/2)$. This means that the total current through each
junction in this case is
\begin{equation}\label{icirvar_J}
i_{1,2} = i_b/2 \mp\cos\left(\frac{i_b}{2}\tau\right).
\end{equation}
Oscillating part of this current induces Shapiro steps in
current-voltage (IV) curve of the junctions, see Fig.~\ref{Fig1}(a).
Total width of the first step is \cite{L} $2J_1(i_{cir}^a/w_{J-})$
(where $i_{cir}^a = 1$ is the amplitude of the circulating current
in the limit $l \rightarrow 0$). Since the middle point of the step
lies approximately at IV-curve of individual junction, the step
starts with the current $\sqrt{(i_b/2)^2+1} - J_1(2/i_b) \approx
i_b/2$ that is approximately equal to constant current applied to
each junction.

Considering an increase of the inductance we note that the junctions
are synchronized in antiphase at $\phi_e = \pi/2$. Thus the SQUID
can be approximately analyzed as linear circuit presented in
Fig.~\ref{Fig1}(b), which is a serial connection of two voltage
generators and two resistors corresponding to two Josephson
junctions, and inductance which couples them. From expression
(\ref{icirvar_J}) we deduce that each generator provides harmonic
voltage with normalized amplitude equal to unity. Amplitude of the
current in such circuit is completely coincides with the amplitude
of the considered circulating current (\ref{icirvar})
\begin{equation}\label{icirvar_J_a}
i_{cir}^a = \frac{1}{\sqrt{1+\frac{l^2w^2_{J-}}{4}}}.
\end{equation}

Decrease of the amplitude of the oscillating current flowing through
each junction with increase of the inductance leads to decrease of
the first Shapiro step width. For fixed constant bias current this,
in turn, leads to decrease of the oscillation frequency that shifts
the Shapiro step, leading to synchronization of the junctions at
lower frequency, see Fig.~\ref{Fig1}(a). This means that the
decrease of the time-averaged SQUID voltage is caused by filtering
of the circulating current, and the junctions are switched at
frequency allowed by $LR$-relaxation time of the circuit.

Returning to expression (\ref{SQwjtot1}) we note that at the point
$\phi_e = \pi/2$ where $\tan^2\psi_-$ is infinite the finite value
of $w_J$ can be found through the limit for the last two factors of
the second term
\begin{equation} \label{limSing}
\lim_{\phi_e \rightarrow \pi/2}\left(\frac{i_b}{2} -
\sqrt{\frac{i_b^2}{4} - \cos^2\phi_e}\right)\tan^2\phi_e \rightarrow
\frac{1}{i_b}.
\end{equation}
Since $w_{J-}(\pi/2) = i_b/2$, the Josephson oscillation frequency
at this point is
\begin{equation}\label{maxV}
w_J(\pi/2) = \frac{i_b}{2}\left[1 - \frac{2 l^2}{l^2 i_b^2 +
16}\right].
\end{equation}
This gives the amplitude of the voltage response $v_{pp} =
w_J(\pi/2) - w_J(0)$
\begin{equation}\label{maxV-minV}
v_{pp} = \frac{i_b}{2}\left[1 - \frac{2 l^2}{l^2 i_b^2 + 16}\right]
- \sqrt{\frac{i_b^2}{4} - 1},
\end{equation}
where the last term should be put zero, if $i_b/2 < 1$.

It is seen that with the bias current increase both frequencies
$w_J(\pi/2),~w_J(0)$ tend to $i_b/2$. Since the voltage response
appears due to excitation of the circulating current, it vanishes in
this case because circulating current portion in the total current
flowing through the junctions becomes small. Since junctions in this
limit are biased high above their critical current, time-averaged
voltage on the SQUID corresponds to voltage drop on the two shunt
resistors connected in parallel.

For the bias current equal to the critical current $i_b = 2$ the
voltage response amplitude decreases with the inductance as
\begin{equation}\label{VppSym}
v_{pp} = 1 - \frac{l^2}{2 l^2 + 8}.
\end{equation}
With inductance increase this amplitude tends to $v_{pp} \rightarrow
0.5$. At the same time, according to (\ref{icirvar_J_a}), the
circulating current amplitude tends to zero for $l \rightarrow
\infty$, and so $v_{pp} \rightarrow 0$. Therefore this dependence of
the amplitude on the inductance $v_{pp}(l)$ (\ref{VppSym}) is
relevant only in the frame of validity of the proposed approach i.e.
for $l \leq 1$.

\subsection{Comparison with numerical calculations}

\begin{figure}[t]
\resizebox{1\columnwidth}{!}{
\includegraphics{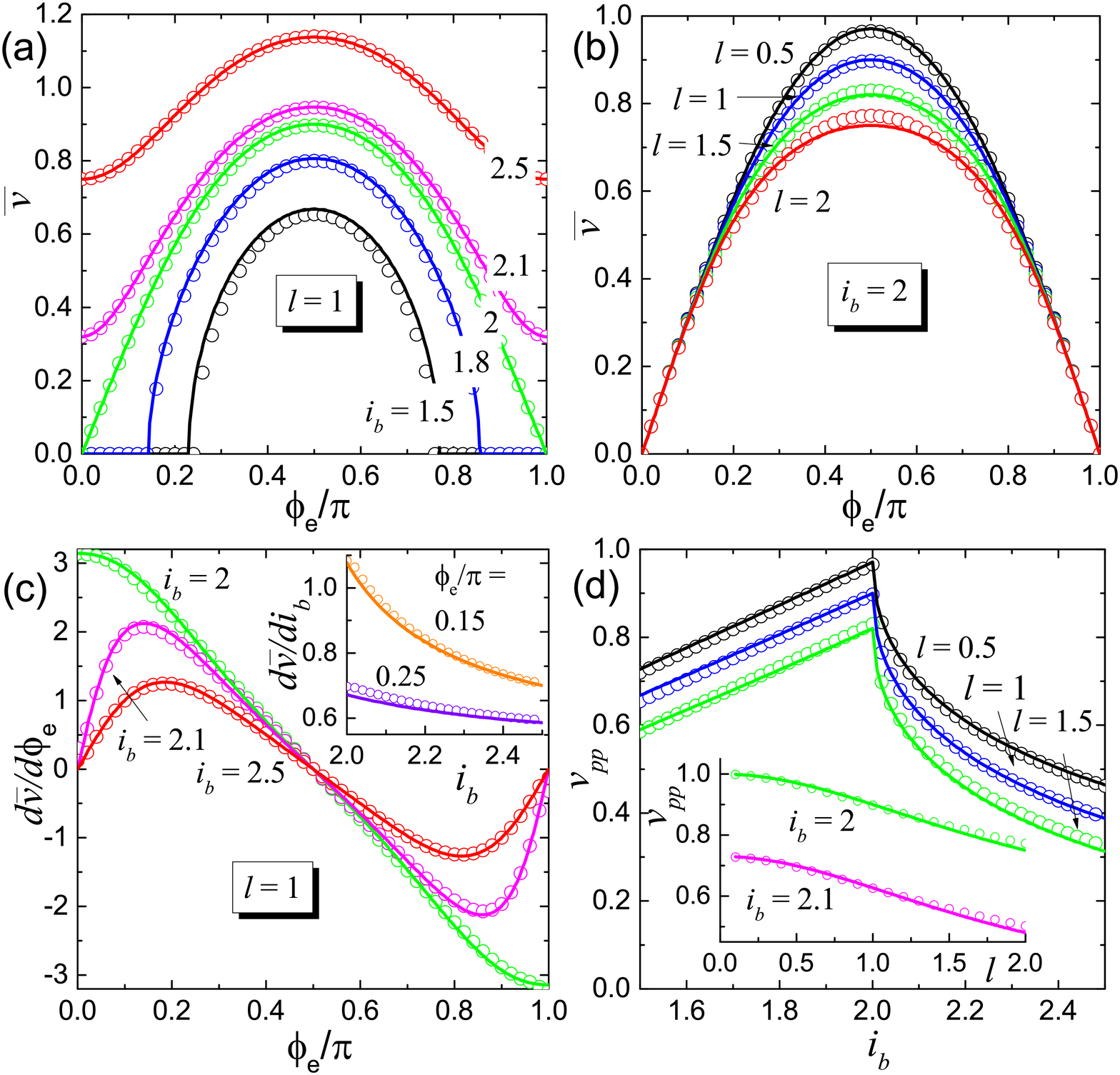}}
\caption{Comparison of results obtained using the presented
analytical approach (expressions (\ref{SQwjtot1}), (\ref{maxV-minV})
- solid curves) and using numerical calculations of system
(\ref{SQUIDeq}) (dots). (a) SQUID voltage response
$\overline{v}(\phi_e)$ at $l = 1$ for a set of the bias currents
$i_b = 1.5, 1.8, 2, 2.1, 2.5$. (b) The voltage response at $i_b = 2$
for a set of the inductance values $l = 0.5, 1, 1.5, 2$. (c) SQUID
transfer function $\partial\overline{v}/\partial\phi_e$ for the bias
current values $i_b = 2, 2.1, 2.5$; inset shows the dynamic
resistance $\partial\overline{v}/\partial i_b$ for the external flux
values $\phi_e = 0.15, 0.25$; $l = 1$. (d) The voltage response
amplitude $v_{pp}$ versus the bias current for the inductance values
$l = 0.5, 1, 1.5$; inset presents the amplitude versus the
inductance at the bias currents $i_b = 2, 2.1$.} \label{Fig3}
\end{figure}

Figure~\ref{Fig3} shows curves calculated using the presented
analytical approach (expressions (\ref{SQwjtot1}), (\ref{maxV-minV})
- solid lines) and obtained using numerical calculations of system
(\ref{SQUIDeq}) (dots).

Figure~\ref{Fig3}(a) presents SQUID voltage responses for the
inductance value $l = 1$. It is seen that the curves calculated
using both analytical and numerical approaches are well consistent
in the wide bias current range $i_b = 1.5 \ldots 2.5$ around the
SQUID critical current.

Equation (\ref{psivar}) shows that amplitude of the difference phase
oscillations is proportional to $\psi^A_\sim \sim -\sin\phi_e$, and
so the inductance mainly affects middle part of the voltage
response. Increase of the inductance leads to increase of
$\psi^A_\sim \sim -li_{cir}^a/2$ and violation of assumption that
$\psi_\sim \ll 1$. This is seen as small deviation of the analytical
curves from the numerical ones in Fig.~\ref{Fig3}(b) for $l > 1$.

We should note that expression (\ref{SQwjtot1}) for the voltage
response allows one to calculate all related curves like the
transfer function $\partial\overline{v}/\partial\phi_e$ or the
dynamic resistance $\partial\overline{v}/\partial i_b$ presented in
figure~\ref{Fig3}(c) and its inset correspondingly.

Fig.~\ref{Fig3}(d) presents the voltage response amplitude
dependencies on the bias current and on the inductance. Comparing
analytical and numerical data we conclude that within the area of
its applicability presented approach describes SQUID voltage
characteristics fairly well.

\section{SQUID current}

This Section is devoted to consideration of symmetrical SQUID
current response. The response is first considered in
superconducting state for arbitrary SQUID inductance. Consideration
of the response in resistive state is conducted for $l \leq 1$ using
expression (\ref{icirvar}) for the circulating current obtained in
the previous Section.

\subsection{Superconducting state}

Time derivatives of the sum and difference phases $\dot{\theta},~
\dot{\psi}$ as well as oscillating part of the difference phase
$\psi_\sim$ in superconducting state is zero. Expression for the
circulating current in this case is given as
\begin{equation}\label{Icir_main}
i_{cir} = \sin\psi\cos\theta,
\end{equation}
which can be combined with (\ref{SQUIDeq}b) to produce
\begin{equation}\label{Icir_sup}
i_{cir} = \sqrt{\cos^2\psi - \frac{i_b^2}{4}}\tan{\psi}.
\end{equation}

This expression can be readily used in the case of vanishing
inductance $l \rightarrow 0$ with $\psi = -\phi_e$. For zero bias
current it simplifies further
\begin{equation}\label{Icir_sup_L0ib0}
i_{cir} = -\sgn(\cos\phi_e)\sin\phi_e.
\end{equation}

For small but nonzero inductances the circulating current was
derived in the frame of perturbation analysis, to first \cite{DLFG}
and second \cite{TDL} order in the inductance. However, for the
inductance value $l \approx 1$ numerical calculation result becomes
inconsistent compared to Ref.~\cite{TDL, DLFG}.

To find the circulating current for arbitrary inductance one has to
solve an equation $f(\psi) = 0$ for transcendental function
\begin{equation}\label{f_trans}
f(\psi) = \frac{l}{2}\sqrt{\cos^2\psi - \frac{i_b^2}{4}}\tan{\psi} +
\psi + \phi_e
\end{equation}
derived from (\ref{SQUIDeq}a), and substitute the found difference
phase into expression (\ref{Icir_sup}). In general case the solution
can be found using the following quasianalytical approach.

The method is based on integral definition of root $x_0$ of
arbitrary continuous function $\mathrm{f}(x)$ having just one zero
in the range of interest $[a,b]$:
\begin{equation}\label{GenApp}
x_0 = a + \int_a^b \Hev[-\sgn(\mathrm{f}[a])] +
\sgn(\mathrm{f}[a])\Hev[\mathrm{f}(x)]dx,
\end{equation}
where $\Hev(x)$ is the Heaviside step function. Since period of
current modulation corresponds to the range $\phi_e \in [0,\pi]$,
root of function $f(\psi)$ has to be sought in the range $\psi_0 \in
[0,-\pi]$. However, function $f(\psi)$ is undefined inside the
region $|\cos\psi| < i_b/2$, and can have up to three zeros
depending on the parameters $i_b$, $l$, $\phi_e$.

\begin{figure}[t]
\begin{center}
\includegraphics[width=6cm]{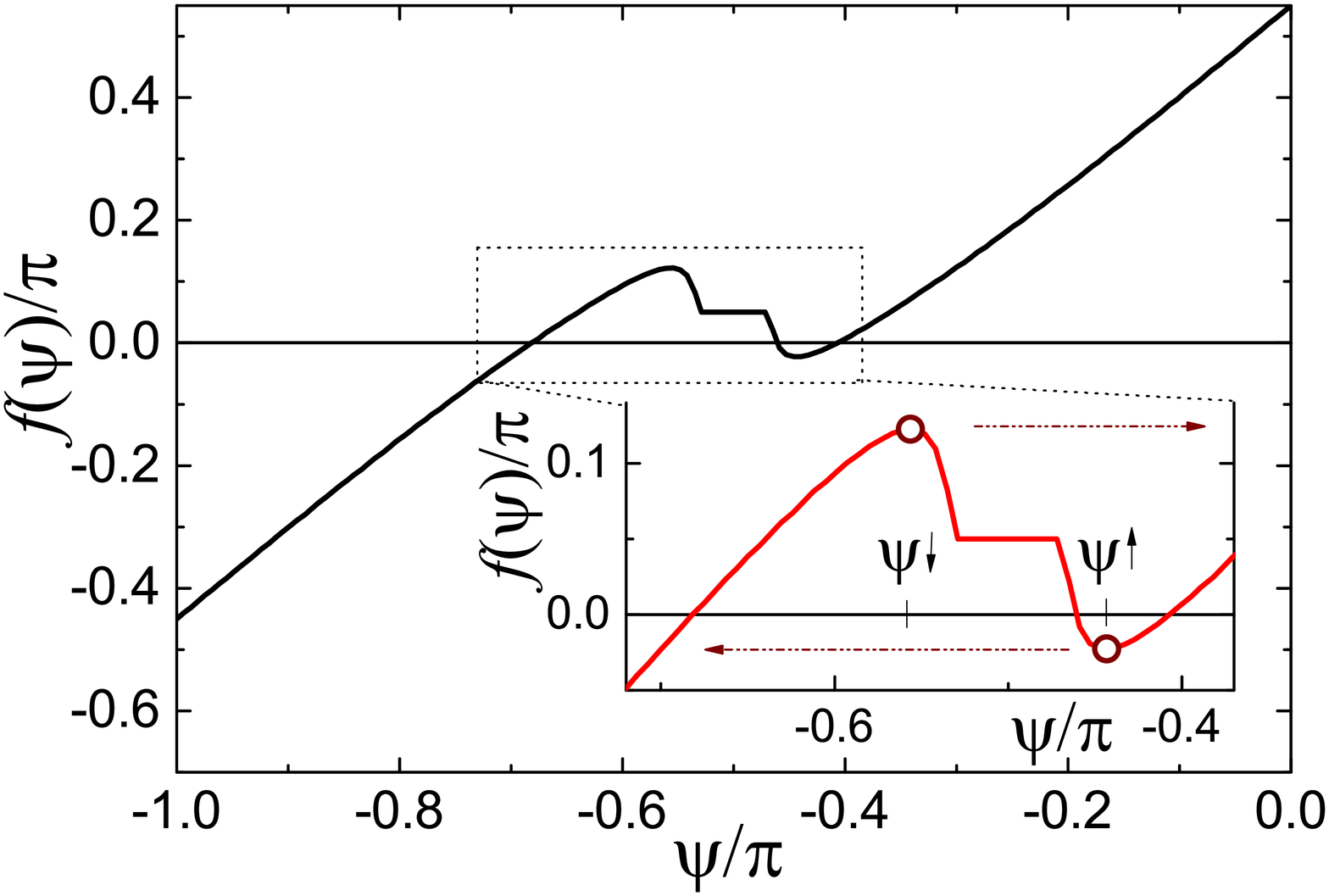}
\end{center}
\caption{Transcendental function $f(\psi)$ (\ref{f_trans})
complementary defined inside the region $|\cos\psi| < i_b/2$, which
implicitly defines the difference phase in superconducting state;
$i_b = 0.2$, $l = 1$, $\phi_e = 0.55\pi$. Inset zooms the function
nonlinear part forming hysteresis at $\psi_0(\phi_e)$ and
$i_{cir}(\phi_e)$ dependencies. Dots show local extrema of $f(\psi)$
function.} \label{Fig4}
\end{figure}

The first obstacle can be overcome by equalizing function $f(\psi)$
to the mean $f_m$ of its boundary values inside the region
$|\cos\psi| \leq i_b/2$, which is obviously $f_m =
\sgn(\psi)\ceil(|\psi|/\pi)\pi/2 + \phi_e$, where $\ceil(x)$ is the
ceiling function. In such complementary definition function
$f(\psi)$ is shown in Fig.~\ref{Fig4}. It is seen from
(\ref{f_trans}) that $\phi_e$ just shifts function $f(\psi)$ along
the ordinate axis.

If function $f(\psi)$ is not monotonic (depending on $l$, $i_b$),
the dependence $\psi_0(\phi_e)$ is hysteretic. In this case the
function can have more than one root for a certain $\phi_e$. To find
proper root, integral (\ref{GenApp}) should be taken from the
starting point $a$ up to the local extremal point closest to $a$
($\psi^\uparrow$ for $\phi_e$ variation $\phi_e^\uparrow =
0\ldots\pi$ and $\psi^\downarrow$ for $\phi_e^\downarrow =
\pi\ldots0$, see inset in Fig.~\ref{Fig4}) if the function $f$ at
these points ($f(a)$ and $f(\psi^{\uparrow\text{or}\downarrow}$) is
of the opposite sign. Otherwise the limits of integration should be
leaved unchanged.

Alteration of the integral upper limit $b$ can be formalized as
follows:
\begin{equation}
b(\phi^{\uparrow\downarrow}_e) = b +
\Hev[-f(a)\mathfrak{f}^{\uparrow\downarrow}(\phi_e)](\psi^{\uparrow\downarrow}
- b),
\end{equation}
where
\begin{equation*}
\mathfrak{f}^{\uparrow\downarrow}(\phi_e) = \cases{
f(\psi^{\uparrow\downarrow};\phi_e) & \text{if~~~$\psi^{\uparrow\downarrow} \neq -\pi/2$},\\
\mp l/2 - \pi/2 + \phi_e & \text{otherwise.}}
\end{equation*}
The reason of such definition of $\mathfrak{f}^{\uparrow\downarrow}$
is the fact that at $i_b = 0$ the function $f$ has a form
\begin{equation}\label{f_trans ib=0}
f(\psi) = \frac{l}{2}\sgn(\cos\psi)\sin\psi + \psi + \phi_e,
\end{equation}
which corresponds to the leap at $\psi^{\uparrow} =
\psi^{\downarrow} = -\pi/2$. Thus we put
$\mathfrak{f}^{\uparrow\downarrow}$ equal to the margin values of
$f$ function in vicinity of the leap.

For $i_b > 0$ coordinates $\psi^\uparrow$, $\psi^\downarrow$ can be
found using derivative of $f(\psi)$ function
\begin{equation}\label{f' trans}
\frac{\partial f}{\partial \psi} =
-\frac{\frac{l}{2}\sin^2\psi}{\sqrt{\cos^2\psi - \frac{i_b^2}{4}}} +
\frac{\frac{l}{2}\sqrt{\cos^2\psi - \frac{i_b^2}{4}}}{\cos^2\psi} +
1
\end{equation}
and formula (\ref{GenApp})
\begin{subequations}
\begin{equation}
\psi^\uparrow = \int_0^{-\arccos\frac{i_b}{2}}\Hev
\left[\frac{\partial f(\phi)}{\partial \phi}\right]d\phi,
\end{equation}
\begin{equation}
\psi^\downarrow = -\pi - \psi^\uparrow.
\end{equation}
\end{subequations}
Note, that $f(\psi^\uparrow) - f(\psi^\downarrow)$ defines the width
of $\psi_0(\phi_e)$ and $i_{cir}(\phi_e)$ hysteresis.

$\psi_0(\phi_e)$ dependencies for both directions of $\phi_e$
variation can be obtained as follows:
\begin{subequations} \label{psi(phie)}
\begin{equation}
\psi_0(\phi_e^\uparrow) = \int_0^{-\pi +
\Hev[-\mathfrak{f}^\uparrow(\phi_e)](\psi^\uparrow +
\pi)}\Hev[f(\phi;\phi_e^\uparrow)]d\phi,
\end{equation}
\begin{equation}
\psi_0(\phi_e^\downarrow) = -\pi +
\int_{-\pi}^{\Hev[\mathfrak{f}^\downarrow(\phi_e)]\psi^\downarrow} 1
- \Hev[f(\phi;\phi_e^\downarrow)]d\phi.
\end{equation}
\end{subequations}

\begin{figure}[t]
\resizebox{1\columnwidth}{!}{
\includegraphics{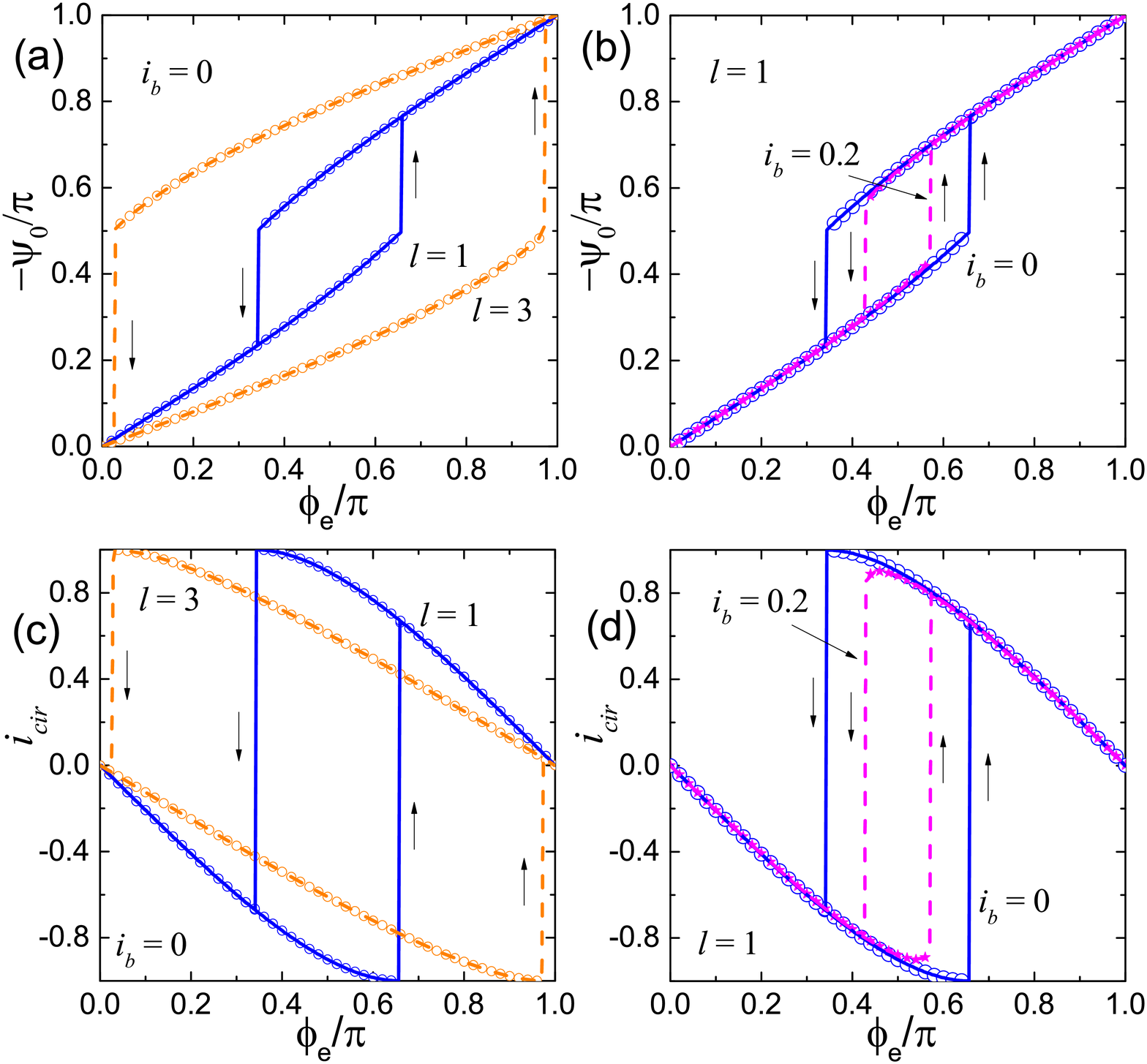}}
\caption{The difference phase $\psi_0$ versus the external flux
$\phi_e$ calculated using expressions (\ref{psi(phie)}) (solid
lines) for (a) the bias current value $i_b = 0$ and the inductance
values $l = 1,~3$, and for (b) $i_b = 0,~0.2$, $l = 1$. (c), (d) The
circulating current corresponding to the curves shown in (a), (b)
panels calculated using expression (\ref{Icir_sup}) (solid lines).
The same dependencies obtained using numerical calculations of
system (\ref{SQUIDeq}) in all panels are presented by dots. Vertical
arrows show direction of variation of the functions.} \label{Fig5}
\end{figure}

These dependencies (\ref{psi(phie)}) and corresponding circulating
currents obtained by expression (\ref{Icir_sup}), as well as
corresponding data obtained by numerical calculations of system
(\ref{SQUIDeq}) are presented in Fig.~\ref{Fig5}. Because of
quasi-analytical nature of the presented approach the data obtained
using expressions (\ref{psi(phie)}), (\ref{Icir_sup}) are perfectly
consistent with the ones calculated numerically.

Increase of the inductance value clearly leads to enlargement of the
hysteresis (see Fig.~\ref{Fig5}(a),(c)) since inductance is the
amplitude of non-monotonic term of $f$ function. Physical meaning of
this fact is as follows. For zero bias current and vanishing
inductance $l \rightarrow 0$ a fluxon penetrates into the SQUID at
$\Phi_e = \Phi_0/2$ that is accompanied by changing of the
circulating current flow direction (\ref{Icir_sup_L0ib0}). The
current flow through a finite inductance produces additional
magnetic flux, and so the external flux can be screened without
penetration of a fluxon into the SQUID loop up to higher values
(Fig.~\ref{Fig5}(c)). The inductance value $l = \pi$ corresponds to
the case where a fluxon penetrates into the SQUID at $\Phi_e =
\Phi_0$. Starting from this inductance value the circuit has two
stable states at zero applied flux (with and without a fluxon inside
it), and thus the loop is called ``quantizing'' one.

At the same time, increase of the bias current leads to widening of
the range where the $f$ function is initially undefined and decrease
of the range where it is non-monotonic. Physically, this means that
the bias current flowing through the junctions allows smaller
circulating current in the superconducting state. This shrinks the
hysteresis (see Fig.~\ref{Fig5}(b),(d)) and finally results in
formation of the region of resistive state.

\subsection{Resistive state}

\begin{figure}[t]
\resizebox{1\columnwidth}{!}{
\includegraphics{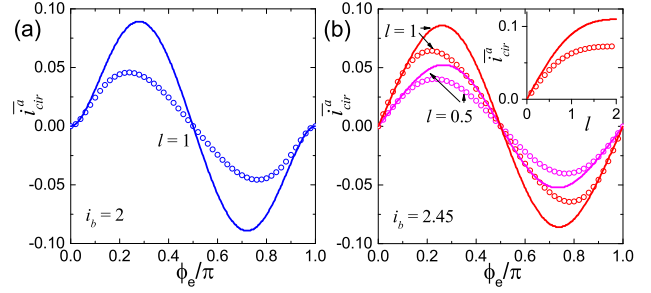}}
\caption{The averaged circulating current $\overline{i}_{cir}$
versus the external flux $\phi_e$ calculated using expression
(\ref{IcirR1fin}) with $\psi_- = -\phi_e$ (solid lines) and obtained
by numerical calculations of system (\ref{SQUIDeq}) (dots). (a) $i_b
= 2$, $l = 1$. (b) $i_b = 2.45$, $l = 0.5, 1$; inset shows the
averaged current at $\phi_e = \pi/4$ versus the inductance
calculated using (\ref{IcirR1_ampl_Sym}) (solid line) and
numerically (dots).} \label{Fig6}
\end{figure}

Time averaging of (\ref{icirvar}) leads to the following expression
for the averaged circulating current in resistive state:
\begin{equation} \label{IcirR1fin}
\overline{i}_{cir} = -\frac{2 l w_{J-}}{l^2 w_{J-}^2 +
4}\left(\frac{i_b}{2} - w_{J-}\right)\tan\psi_-.
\end{equation}
Taking $\psi_- = -\phi_e$ we calculated current curves presented in
Fig.~\ref{Fig6} by solid lines. Corresponding curves calculated
numerically using system (\ref{SQUIDeq}) are shown by dots.

Zero value of the averaged current at $\phi_e = 0, ~\pi$ corresponds
to an absent or purely harmonic circulating current, according to
(\ref{icirvar}). Nonzero averaged current at $\phi_e \approx \pi/4$
appears due to unharmonic shape of the circulating current which is
not accurately described with proposed linearization of the
equations (\ref{SQUIDeq}). While consistency for the bias current
$i_b = 2$ is only qualitative (Fig.~\ref{Fig6}(a)), for higher bias
current value it is improved (Fig.~\ref{Fig6}(b)). The reason is
decrease of oscillating part of the difference phase $\psi_\sim$
(\ref{psivar}).

Simplifying expression (\ref{IcirR1fin}) at the point $\phi_e =
\pi/4$, one can estimate the averaged circulating current amplitude
$\overline{i}^{~a}_{cir}$,
\begin{equation} \label{IcirR1_ampl}
\overline{i}^{~a}_{cir} = \frac{2 l \sqrt{i_b^2 - 2}\left(i_b -
\sqrt{i_b^2 - 2}\right)}{l^2\left(i_b^2 - 2\right) + 16}.
\end{equation}
For the bias current value $i_b = \sqrt{6} \approx 2.45$ the
amplitude is approximately
\begin{equation} \label{IcirR1_ampl_Sym}
\overline{i}^{~a}_{cir} \approx \frac{0.45 l}{l^2 + 4}.
\end{equation}
Corresponding curve $\overline{i}^{~a}_{cir}(l)$ is shown in inset
of Fig.~\ref{Fig6}(b) with numerical data for comparison.

\section{SQUID with small spread of parameters}

In this Section we study effect of reasonable technological spread
of parameters ($\Delta I_c,\Delta R_n$ up to $\pm 20\%$, and
arbitrary inequality of SQUID inductive shoulders) on the SQUID
responses. Such spread may result in a small asymmetry of the SQUID.
\begin{figure}[t]
\begin{center}
\includegraphics[width=4.5cm]{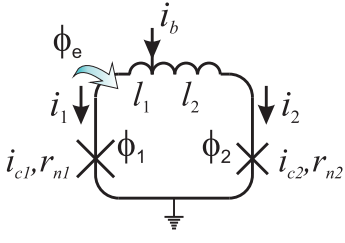}
\end{center}
\caption{DC SQUID with asymmetry presented by inequality of the
inductances $l_1 \neq l_2$, critical currents $i_{c1} \neq i_{c2}$,
and shunt resistances $r_{n1} \neq r_{n2}$.} \label{Fig7}
\end{figure}

The considered DC SQUID is shown in Fig.~\ref{Fig7}. The asymmetry
is presented by inequality of the inductive shoulders $l_1 \neq
l_2$, $l_1 + l_2 = l$, and differences of the critical currents and
shunt resistances $i_{c1} \neq i_{c2}$, $r_{n1} \neq r_{n2}$ of the
junctions. We assume below that $|i_{c1} - i_{c2}| \ll 1$, $|r_{n1}
- r_{n2}| \ll 1$, and that the bias current is of the order of the
SQUID critical current $i_b \approx i_{c1} + i_{c2}$.

\subsubsection{Asymmetry of the inductive shoulders.}

Simplest case of SQUID asymmetry is inequality of the inductances
$l_1 \neq l_2$. If $i_{c1} = i_{c2}$, $r_{n1} = r_{n2}$ then
modification appears only in equation (\ref{SQUIDeq}a). Phase drop
on the inductances $l_1 i_1 - l_2 i_2$ rewritten as $(i_b/2)[l_1 -
l_2] + (l/2)[i_1 - i_2]$ converts this equation into the following
one:
\begin{equation}
\frac{\Sigma l}{2}\dot{\psi} = -\left(\psi + \phi_e +
\frac{i_b}{2}\frac{\Delta l}{2}\right) - \frac{\Sigma
l}{2}\sin\psi\cos\theta,
\end{equation}
where $\Sigma l = l_1 + l_2$, $\Delta l = l_1 - l_2$.

It is seen that this asymmetry can be accounted just by the constant
offset $i_b\Delta l/4$ to the external flux $\phi_e$, which appears
due to the bias current flow through the part of inductance $\Delta
l$. In consideration of asymmetries of the critical currents and
shunt resistances we first put $\Delta l = 0$ for simplicity, and
then generalize obtained results for arbitrary $\Delta l$.

\subsubsection{Asymmetry of the critical currents and shunt resistances.}

The spread in critical currents of Josephson junctions and shunt
resistances affects system (\ref{SQUIDeq}) noticeably, and these
equations can be rewritten as
\begin{subequations} \label{asymSQeq}
\begin{multline}
\frac{l}{\Sigma r_n}\dot{\psi} = -(\psi + \phi_e) +
\frac{l}{2}\frac{i_b}{2}\frac{\Delta r_n}{\Sigma r_n}\\
-\frac{l}{\Sigma r_n}\left(\frac{\Sigma v_c}{2}\sin\psi\cos\theta +
\frac{\Delta v_c}{2}\cos\psi\sin\theta\right),
\end{multline}
\begin{multline}
\dot{\theta} =  \frac{i_b}{2}\frac{\Sigma r_n}{2} - \frac{\psi +
\phi_e}{l}\Delta r_n\\ -\frac{\Sigma v_c}{2}\cos\psi\sin\theta -
\frac{\Delta v_c}{2}\sin\psi\cos\theta.
\end{multline}
\end{subequations}
Here $v_c = i_c r_n$ is the characteristic voltage, $\Sigma v_c =
v_{c1} + v_{c2}$, $\Delta v_c = v_{c1} - v_{c2}$, $\Sigma r_n =
r_{n1} + r_{n2}$, $\Delta r_n = r_{n1} - r_{n2}$.

Below we first consider influence of this asymmetry on the voltage
response and then describe its effect on the circulating current.
The difference phase is again presented as a sum of constant
$\psi^*_-$ and oscillating $\psi^*_\sim$ parts, where the last one
is assumed to be small $\psi^*_\sim \ll 1$.

\subsection{Voltage on SQUID with small asymmetry}

\subsubsection{Solution for $\psi = \psi^*_-$.}

Even in the absence of the asymmetry of the inductive shoulders
asymmetry of the critical currents and shunt resistances still leads
to a constant offset of the external flux. This offset can be found
by time averaging of equation (\ref{asymSQeq}a) at $\phi_e = 0$ and
$i_b \approx 2$, assuming that $i_{c1,2},r_{n1,2}\approx 1$. Since
$\overline{\dot{\psi}}_- = 0$, $\overline{\cos\theta} = 0$, and
$\overline{\sin\theta} = [(i_b/2) - w_{J-}]/\cos\psi_- \approx 1$,
under the made assumptions the difference phase is
\begin{equation} \label{Psi0*}
\psi_-^* = -\phi_e -\frac{l}{2}\left[\frac{\Delta i_c}{2} -\frac{i_b
- \Sigma i_c}{2}\frac{\Delta r_n}{\Sigma r_n}\right],
\end{equation}
where $\Sigma i_c = i_{c1} + i_{c2}$, $\Delta i_c = i_{c1} -
i_{c2}$.

The asymmetries $\Delta i_c$ and $\Delta r_n$ provide redistribution
of the bias current toward the junction with larger $i_c$ but
smaller $r_n$. Corresponding terms enter in (\ref{Psi0*}) with
opposite signs accordingly. Effect due to asymmetry of the shunt
resistances refers to resistive state that is manifested by
prefactor proportional to the bias current deviation from the SQUID
critical current $i_b - \Sigma i_c$.

To find the sum phase it is convenient to present equation
(\ref{asymSQeq}b) in the following form:
\begin{multline} \label{Theta0*eq}
\dot{\theta} =  \frac{i_b}{2}\frac{\Sigma r_{n}}{2} - \frac{\psi +
\phi_e}{l}\Delta r_{n}\\ -\sqrt{\frac{\Delta v_{c}^2}{4} + v_{c1}v_{c2}\cos^2\psi}\\
\times\sin\left(\theta + \arctan\left[\frac{\Delta v_c}{\Sigma
v_c}\tan\psi\right]\right).
\end{multline}
Structure of this equation is quite analogous to the one of equation
(\ref{SQUIDeq}b), and so its solution replicates equation
(\ref{SolFor0ap}). The only difference here is the shift of the sum
phase:
\begin{equation} \label{SolFor0ap*}
\tan\frac{\theta+c}{2} = z^* + \sqrt{1
-z^{*2}}\tan\frac{\frac{i_b^*}{2}\sqrt{1-z^{*2}}\tau}{2},
\end{equation}
where
\begin{equation*}
\frac{i_b^*}{2} = \frac{i_b}{2}\frac{\Sigma r_{n}}{2} +
\left[\frac{\Delta i_c}{2} - \frac{i_b - \Sigma i_c}{2}\frac{\Delta
r_n}{\Sigma r_n}\right]\frac{\Delta r_n}{2},
\end{equation*}
\begin{equation*}
z^* = \frac{2}{i_b^*}\sqrt{\frac{\Delta v_c^2}{4} +
v_{c1}v_{c2}\cos^2\psi_-^*},
\end{equation*}
\begin{equation*}
c = \arctan\left[\frac{\Delta v_c}{\Sigma v_c}\tan\psi_-^*\right].
\end{equation*}

The Josephson frequency similarly follows from equation
(\ref{SolFor0ap*}):
\begin{multline} \label{wJ0*}
w_{J-}^* = \sqrt{v_{c1}v_{c2}} \\ \times\sqrt{\frac{[(i_b -
i_{c1})r_{n1} + v_{c2}][(i_b - i_{c2})r_{n2} +
v_{c1}]}{i_{c1}i_{c2}\Sigma r_{n}^2} - \cos^2\psi_-^*},
\end{multline}
where $w_{J-}^*$ should be put zero if the expression under the
second square root is negative.

Here the square root of the characteristic voltage product
$v_{c1}v_{c2}$ is a natural scaling factor for the voltage.
Presenting this factor as
\begin{equation}\label{Vc1Vc2}
\sqrt{v_{c1}v_{c2}} = \frac{\sqrt{(\Sigma i_c^2 - \Delta
i_c^2)(\Sigma r_n^2 - \Delta r_n^2)}}{4},
\end{equation}
we note that the differences of the critical currents and shunt
resistances provide similar contributions.

Considering the first term under the second square root of
expression (\ref{wJ0*}) as squared effective bias current divided by
the squared SQUID critical current
\begin{equation}
\frac{i_{b_{eff}}^2}{\Sigma i_c^2} = \frac{[(i_b - i_{c1})r_{n1} +
v_{c2}][(i_b - i_{c2})r_{n2} + v_{c1}]}{i_{c1}i_{c2}\Sigma r_{n}^2},
\end{equation}
by analogy with expression (\ref{SQwj0}), one can study how this
ratio differs from the one for symmetrical case.

For identification of effect of the critical current difference on
this ratio it is convenient to put difference of the shunt
resistances equal to zero $\Delta r_n = 0$ and $\Sigma r_n = 2$. In
this case the considered ratio is
\begin{equation}\label{IbeffDic_orig}
\frac{i_{b_{eff}}^2}{\Sigma i_c^2} = \frac{i_b^2 - \Delta
i_c^2}{\Sigma i_c^2 - \Delta i_c^2},
\end{equation}
and the squared effective bias current is
\begin{equation}\label{IbeffDic}
i_{b_{eff}}^2 = i_b^2 + \frac{\Delta i_c^2}{\Sigma i_c^2 - \Delta
i_c^2}\left(i_b^2 - \Sigma i_c^2 \right).
\end{equation}

It is seen that asymmetry of the critical currents $\Delta i_c$
affects both $i_b$ and $\Sigma i_c$ (\ref{IbeffDic_orig}). This
results in tiny deviation of $i_{b_{eff}}^2$ from $i_b^2$
proportional to the asymmetry in square, and multiplied by deviation
of the squares of the bias current from the SQUID critical current
(\ref{IbeffDic}).

Consideration of shunt resistance difference at $\Delta i_c = 0$,
$\Sigma i_c = 2$ leads to the following expression
\begin{equation}\label{IbeffDrn}
\frac{i_{b_{eff}}^2}{4} = \frac{i_b^2}{4} - \left(\frac{i_b -
2}{2}\right)^2\frac{\Delta r_n^2}{\Sigma r_n^2} .
\end{equation}

Difference of the shunt resistances affects the effective bias
current in accordance with the bias current redistribution (see also
(\ref{Psi0*})). In comparison with the effect provided by difference
of the critical currents, here the deviation of $i_{b_{eff}}^2$ from
$i_b^2$ is even smaller (since $i_b^2 - \Sigma i_c^2 > (i_b - \Sigma
i_c)^2$ for $i_b
> i_c$, see (\ref{IbeffDic}), (\ref{IbeffDrn})).

Obtained results (\ref{Psi0*}), (\ref{IbeffDic}), (\ref{IbeffDrn})
allow us to conclude that at $i_b \approx \Sigma i_c$ the Josephson
oscillation frequency is approximately equal to
\begin{equation} \label{wJ0*ib=ic}
w_{J0}^* \approx \sqrt{v_{c1}v_{c2}}\sqrt{1 - \cos^2\left(\phi_e +
\frac{l}{2}\frac{\Delta i_c}{2}\right)},
\end{equation}
and the asymmetries $\Delta i_c$, $\Delta r_n$ mainly scale the
averaged voltage and provide offset to the external flux
corresponding to $\Delta i_c$.

\begin{figure}[t]
\resizebox{1\columnwidth}{!}{
\includegraphics{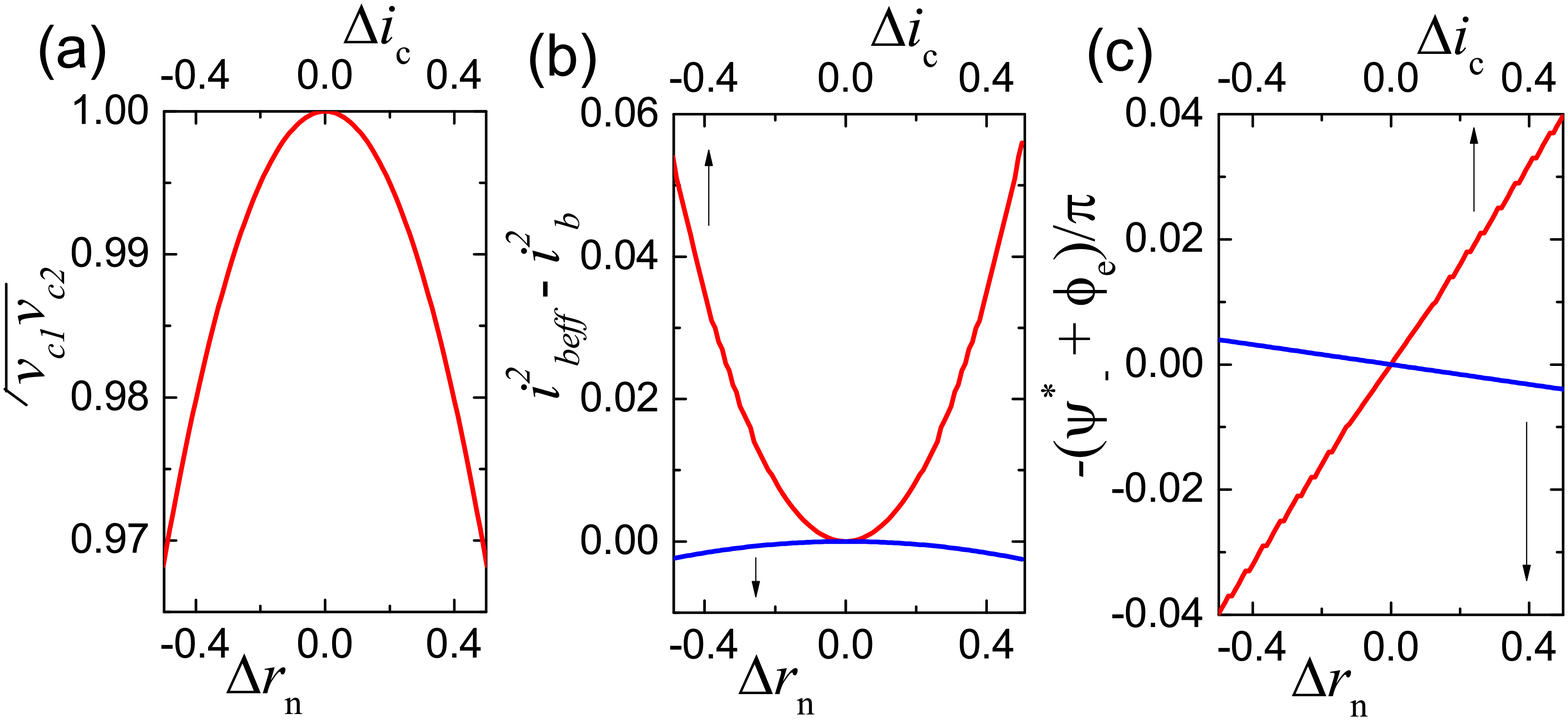}}
\caption{(a) Scaling factor $\sqrt{v_{c1}v_{c2}}$ for the averaged
voltage (\ref{wJ0*}), (b) deviation of the squared effective bias
current from the squared bias current $i_{b_{eff}}^2 - i_b^2$, and
(c) offset to the bias flux $-(\psi^*_- + \phi_e)/\pi$ versus
$\Delta r_n$ at $\Delta i_c = 0$, and versus $\Delta i_c$ at $\Delta
r_n = 0$; $\Sigma i_c, \Sigma r_n = 2$, $i_b = 2.2$.} \label{Fig8}
\end{figure}

Considered effects of asymmetries of the critical currents and shunt
resistances on the scaling factor $\sqrt{v_{c1}v_{c2}}$, the
deviation of $i_{b_{eff}}^2$ from $i_b^2$, and the bias flux offset
are illustrated in Fig.~\ref{Fig8}.

\subsubsection{Solution for $\psi = \psi^*_- + \psi^*_\sim$.}

Impact of oscillating current circulating in the SQUID can be
introduced by oscillating part of the difference phase. Substitution
of $\psi = \psi_-^* + \psi_\sim^*$ into equation (\ref{asymSQeq}a)
leads to
\begin{multline}\label{Psi1*Eq}
\frac{l}{\Sigma r_n}\dot{\psi}^*_\sim = -\psi^*_\sim +
\frac{l}{2}\left[\frac{\Delta i_c}{2}
+\frac{\Sigma i_c}{2}\frac{\Delta r_n}{\Sigma r_n}\right] \\
-\frac{l}{\Sigma r_n}\left(\frac{\Sigma
v_c}{2}\sin\psi_-^*\cos\theta + \frac{\Delta
v_c}{2}\cos\psi_-^*\sin\theta\right).
\end{multline}
It is hard to derive exact analytical solution of this equation,
while approximate solution can be drawn rather easy assuming $l \ll
1$ and putting $\dot{\theta} \approx \overline{\dot{\theta}} =
w^*_{J-}$:
\begin{multline}\label{Psi1*}
\psi_\sim^* = \frac{l}{2}\left[\frac{\Delta i_c}{2} +\frac{\Sigma
i_c}{2}\frac{\Delta r_n}{\Sigma r_n}\right]-\frac{l}{\Sigma
r_n\sqrt{1 + \left[\frac{lw^*_{J-}}{\Sigma
r_n}\right]^2}}\\\times\bigg[\frac{\Sigma
v_c}{2}\sin\psi_-^*\cos\left(\theta-\arctan\left[\frac{lw^*_{J-}}{\Sigma
r_n}\right]\right)\\+\frac{\Delta
v_c}{2}\cos\psi_-^*\sin\left(\theta-\arctan
\left[\frac{lw^*_{J-}}{\Sigma r_n}\right]\right)\bigg].
\end{multline}

To find the correction $w^*_{J\sim}$ of the Josephson frequency we
substitute the found difference phase $\psi = \psi^*_- +
\psi^*_\sim$ (\ref{Psi0*}), (\ref{Psi1*}) into equation
(\ref{asymSQeq}b):
\begin{multline} \label{wJ1*e}
w_{J\sim}^* = -\frac{\overline{\psi_\sim^*}}{l}\Delta r_n\\ +
\frac{\Sigma v_c}{2}\sin\psi_-^*(\overline{\psi_\sim^*\sin\theta}) -
\frac{\Delta v_c}{2}\cos\psi_-^*(\overline{\psi_\sim^*\cos\theta}).
\end{multline}
Since $\psi^*_\sim$ is proportional to the inductance (\ref{Psi1*}),
the last two terms in (\ref{wJ1*e}) represent SQUID $LR$-filtering
of the circulating current. They vanish in the limit $l \rightarrow
0$. At the same time, for $l = 0$ the first term is non zero
representing difference of voltage drops of the time-averaged
circulating current on the shunt resistors.

Unfortunately, full expression for $w_{J\sim}^*$ is rather
complicated
\begin{multline} \label{wJ1*}
w_{J\sim}^* = \\ -\frac{\frac{l}{\Sigma
r_n}\frac{i_b^*}{2}\left(\frac{l w_{J-}^*}{\Sigma r_n} v_{c1}^2
v_{c2}^2 \sin^2 2\psi_-^* + K_1 + K_2\right)}{4\left(\left[\frac{l
w_{J-}^*}{\Sigma r_n}\right]^2 + 1\right)\left[\frac{i_b^{*2}}{4} -
w_{J-}^{*2}\right]\left(\frac{i_b^*}{2} + w_{J-}^* \right)} + K_3.
\end{multline}
Here the coefficients $K_{1,2,3}$ have following forms
\begin{multline*}
K_1 = \frac{v_{c1}v_{c2}}{i_b^*}\left[\Sigma v_c \Delta v_c - 4
w_{J-}^* \frac{\Delta r_n}{\Sigma r_n}\left(\frac{i_b^*}{2} +
w_{J-}^* \right)\right]\\
\times\left(\frac{i_b^*}{2} - w_{J-}^* \right)\sin 2\psi_-^*,
\end{multline*}
\begin{multline*}
K_2 = \frac{\Sigma v_c \Delta
v_c}{i_b^*}\bigg[\frac{w_{J-}^{*2}}{2}\frac{l}{\Sigma r_n}\Sigma v_c
\Delta v_c\\ - 2\frac{\Delta r_n}{l}\left(\frac{i_b^{*2}}{4} -
w_{J-}^{*2}\right) \bigg],
\end{multline*}
\begin{multline*}
K_3 = \\ \frac{l}{2}\left(\frac{\Delta i_c}{2} +\frac{\Sigma
i_c}{2}\frac{\Delta r_n}{\Sigma
r_n}\right)\left[\frac{v_{c1}v_{c2}}{2}\frac{\sin
2\psi_-^*}{\left(\frac{i_b^*}{2} + w_{J-}^* \right)} - \frac{\Delta
r_n}{l} \right].
\end{multline*}
They arise due to asymmetry of the SQUID parameters.

Amplitude of the voltage response can be found through the total
Josephson frequency $w_J^* = w_{J-}^* + w_{J\sim}^*$ taken at
$\psi_-^* = -\pi/2$ and $\psi_-^* = 0$. For the bias current equal
to the SQUID critical current $i_b = \Sigma i_c$ this amplitude is
\begin{multline} \label{VppAsym}
v^*_{pp} = \sqrt{v_{c1}v_{c2}}-\frac{\Sigma v_c}{4}\\ \times
\left[\frac{v_{c1}v_{c2}\frac{l^2}{\Sigma r_n^2}\Sigma v_c -
\frac{\Delta r_n}{\Sigma r_n} \Delta v_c}{\left(\sqrt{v_{c1}v_{c2}}
+ \frac{\Sigma v_c}{2}\right)\left[v_{c1}v_{c2}\frac{l^2}{\Sigma
r_n^2} + 1\right]} + \frac{2 \Delta r_n \Delta v_c}{\Sigma r_n\Sigma
v_c} \right].
\end{multline}
For symmetrical case this expression takes the form of
(\ref{VppSym}) with according scaling coefficients.

Note, that for $l = 0$ this amplitude (\ref{VppAsym}) is
\begin{multline} \label{VppAsymL0}
v^*_{pp} = \sqrt{v_{c1}v_{c2}}-\frac{\Delta v_c \Delta r_n}{2 \Sigma
r_n}\left(1 - \frac{\Sigma v_c}{\Sigma v_c + 2\sqrt{v_{c1}v_{c2}}}
\right).
\end{multline}
The last term in (\ref{VppAsymL0}) represents deviation of the
voltage response amplitude corresponding to the first term of
(\ref{wJ1*e}). For zero difference of the shunt resistances $\Delta
r_n = 0$ this term is zero. At the same time, for zero difference of
the critical currents $\Delta i_c = 0$ ($\Delta v_c = \Sigma i_c
\Delta r_n/2$) this deviation is proportional to $\Delta r_n^2$. In
this case the difference of shunt resistances always leads to
additional decrease of the voltage response. We will return to
consideration of this fact below in Section devoted to time-averaged
circulating current.

\subsubsection{Generalization to the case of unequal inductive
shoulders.}

Since asymmetry of the inductances $l_1 \neq l_2$ just shifts the
bias flux, it can be accounted by according addition $i_b \Delta
l/4$ to the expression for the constant difference phase
(\ref{Psi0*})
\begin{equation} \label{Psi0*total}
\psi_-^* = -\phi_e -\frac{i_b}{2}\frac{\Delta l}{2} -\frac{\Sigma
l}{2}\left[\frac{\Delta i_c}{2} -\frac{i_b - \Sigma
i_c}{2}\frac{\Delta r_n}{\Sigma r_n}\right],
\end{equation}
and substitution of the total inductance $\Sigma l$ instead of $l$
in equation (\ref{Psi1*Eq}) and subsequent expressions
(\ref{Psi1*})-(\ref{VppAsym}).

\begin{figure}[t]
\resizebox{1\columnwidth}{!}{
\includegraphics{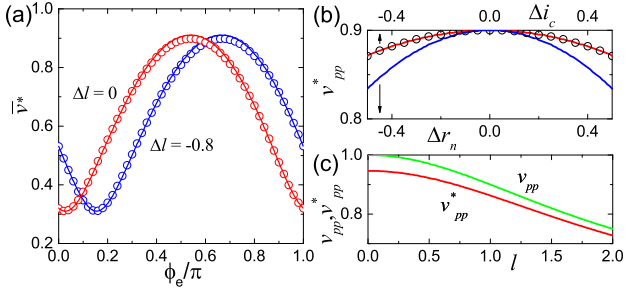}}
\caption{(a) Voltage response calculated using expression (\ref{v*})
(solid lines) and calculated numerically using system
(\ref{asymSQeq}) (dots) for a set of parameters: $i_b = 2$, $\Sigma
i_c = 1.9$, $\Delta i_c = -0.3$, $\Sigma r_n = 2.05$, $\Delta r_n =
0.35$, $\Sigma l = 1$, and $\Delta l = 0,~-0.8$. (b) Voltage
response amplitude $v^*_{pp}$ versus $\Delta i_c$ at $\Delta r_n =
0$, and versus $\Delta r_n$ at $\Delta i_c = 0$ (solid lines) for
$i_b, \Sigma i_c, \Sigma r_n = 2$ and $l = 1$. Approximate
dependence $v^*_{pp} = v_{pp}\sqrt{v_{c1}v_{c2}}$ derived from
(\ref{VppSym}), (\ref{Vc1Vc2}) is shown by dots. (c) Voltage
response amplitude $v^*_{pp}$ for the same set of parameters as the
ones taken for panel (a) but $i_b = \Sigma i_c$ calculated using
expression (\ref{VppAsym}), and the amplitude $v_{pp}$ for
symmetrical case at $i_b = 2$ calculated using expression
(\ref{VppSym}) versus the inductance.} \label{Fig9}
\end{figure}

Voltage response
\begin{equation} \label{v*}
\overline{v}^* = w_J^* = w_{J-}^* + w_{J\sim}^*
\end{equation}
(where $w_J^*$ should be put zero if $w_{J-}^* = 0$) can be found by
combining expressions (\ref{wJ0*}), (\ref{wJ1*}), and
(\ref{Psi0*total}). The responses calculated from these expressions
for a chosen set of parameters $i_b = 2$, $\Sigma i_c = 1.9$,
$\Delta i_c = -0.3$, $\Sigma r_n = 2.05$, $\Delta r_n = 0.35$,
$\Sigma l = 1$, and $\Delta l = 0,~-0.8$ are shown in
Fig.~\ref{Fig9}(a) by solid lines, while the data obtained by
numerical calculations of system (\ref{asymSQeq}) are presented by
dots. It is seen that the data are well consistent despite the fact
that oscillating part of the difference phase (\ref{Psi1*}) is found
approximately. Even for $\Delta l = 0$ the voltage response has
small constant shift along $\phi_e$ axis due to $\Delta i_c$,
$\Delta r_n$ asymmetries, in accordance with (\ref{Psi0*total}).
Corresponding terms arising in (\ref{wJ1*e}) provide some asymmetry
of the voltage response shape which is though quite small for
considered case $|\Delta i_c|, |\Delta r_n| \ll 1$.

Figure~\ref{Fig9}(b) illustrates effects of the critical current
difference $\Delta i_c$ at $\Delta r_n = 0$ and difference of the
shunt resistances $\Delta r_n$ at $\Delta i_c = 0$ on the amplitude
of the voltage response $v^*_{pp}$ for $i_b, \Sigma i_c, \Sigma r_n
= 2$ and $l = 1$. Corresponding curves are calculated using
(\ref{VppAsym}) and shown by lines. The dependence $v^*_{pp} =
v_{pp}\sqrt{v_{c1}v_{c2}}$, which can be derived from the amplitude
$v_{pp}$ in symmetrical case (\ref{VppSym}) and the voltage scaling
factor (\ref{Vc1Vc2}) is presented by dots.

Consistency of the curves indicates that individual effect of the
critical current difference reduces to the voltage scaling. For
small inductance values ($l \ll 1$) expression (\ref{VppAsym}) can
be approximated by
\begin{equation} \label{v*dic_appr}
v^*_{pp} \approx \sqrt{1-\frac{\Delta i_c^2}{4}}v_{pp}.
\end{equation}

At the same time, individual effect of difference of the shunt
resistances includes additional decrease of $v^*_{pp}$ due to
time-averaged voltage drop of the circulating current on $\Delta
r_n$ that leads to
\begin{equation}\label{v*drn_appr}
v^*_{pp} \approx \sqrt{1-\frac{\Delta r_n^2}{4}}v_{pp} -
\frac{\Delta r_n^2}{4}\frac{l^2+2}{l^2+4}.
\end{equation}

Note, that decrease of the amplitude $v^*_{pp}$ resulting from
$LR$-filtering not precisely follows $v_{pp}$ (\ref{VppSym}).
Therefore obtained expressions (\ref{v*dic_appr}),
(\ref{v*drn_appr}) are relevant only for $l \ll 1$. In general, the
circulating current-time dependence, which is purely harmonic in
symmetrical case at $\psi_- = -\pi/2$ is unharmonic at $\psi^*_- =
-\pi/2$ for $\Delta v_c \neq 0$. This can be seen from expressions
(\ref{SolFor0ap*}), (\ref{Psi1*}). The current-time dependence
becomes inclined (at $\psi^*_- = -\pi/2$) due to different rates of
switching of the junctions $v_{c1} \neq v_{c2}$.
Figure~\ref{Fig9}(c) presents the voltage response amplitude
$v_{pp}^*(l)$ calculated using expression (\ref{VppAsym}) for the
same parameters as the ones taken for Fig.~\ref{Fig9}(a) but $i_b =
\Sigma i_c$. It is compared with the amplitude $v_{pp}$ of
symmetrical SQUID (\ref{VppSym}) versus the total inductance.
Inconstancy of the difference $v_{pp}^* - v_{pp}$ illustrates slight
changes in the circulating current and conditions of its filtering
provided by asymmetry.

\subsection{Circulating current in SQUID with small asymmetry}

\subsubsection{Superconducting state.}

Effect of the asymmetries on the circulating current in
superconducting state $\dot{\theta}, \dot{\psi} = 0$ can be easily
found for $i_b = 0$ by combining equations (\ref{asymSQeq}). Since
for zero bias current inequality of the inductive shoulders plays no
role, we consider here only effects of differences of the critical
currents and shunt resistances. The $f(\psi)$ function (\ref{f_trans
ib=0}) obtained from the system (\ref{asymSQeq}) has the following
form:
\begin{multline} \label{f trans asym}
f^*(\psi) = g(\psi)\frac{l}{2}\sgn(\cos\psi)\sin\psi + \psi +
\phi_e,
\end{multline}
where
\begin{equation}
g(\psi) = \frac{\Sigma v_c + \Delta v_c A_I}{\Sigma r_n \sqrt{1 +
A_I^2\tan^2\psi}}
\end{equation}
is factor arising from asymmetry, and
\begin{equation}
A_I = \frac{\Sigma v_c\Delta r_n - \Delta v_c\Sigma r_n}{\Sigma
v_c\Sigma r_n - \Delta r_n^2}.
\end{equation}

\begin{figure}[t]
\resizebox{1\columnwidth}{!}{
\includegraphics{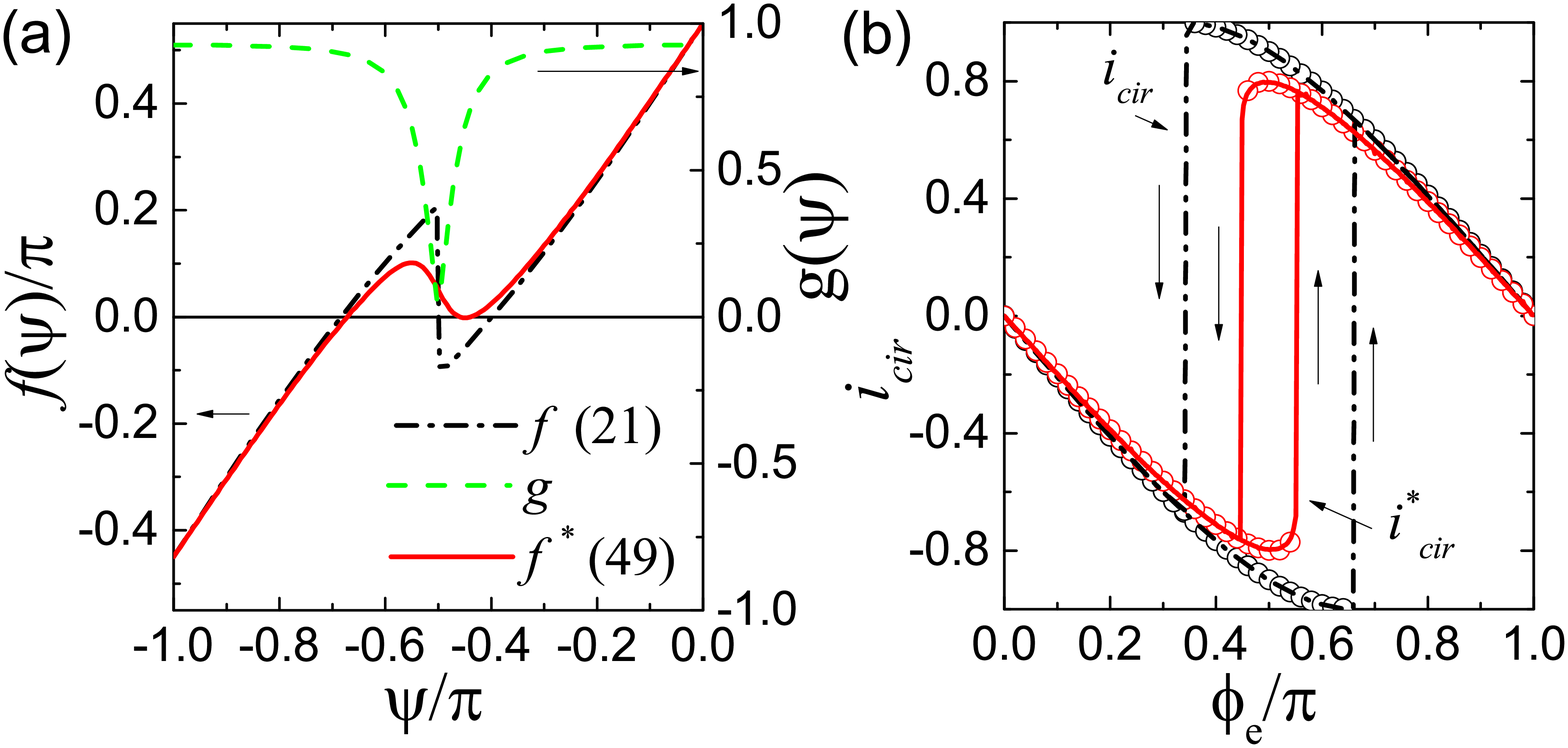}}
\caption{(a) Transcendental function $f$ for symmetrical SQUID
(\ref{f_trans ib=0}), $g(\psi)$ factor affecting its nonlinear term
in asymmetrical case, and $f^*(\psi)$ function (\ref{f trans asym})
for asymmetrical SQUID with parameters $\Sigma i_c = 1.9$, $\Delta
i_c = -0.3$, $\Sigma r_n = 2.05$, $\Delta r_n = 0.35$, $l = 1$. (b)
The circulating currents $i_{cir}$, $i^*_{cir}$ corresponding to the
functions $f$, $f^*$ presented in panel (a), calculated using
expressions (\ref{Icir_sup}), (\ref{psi(phie)}), (\ref{f' trans
asym}) (lines). Numerical data obtained using system
(\ref{asymSQeq}) are shown by dots for comparison. Vertical arrows
show direction of variation of functions.} \label{Fig10}
\end{figure}

Constant $A_I$ is always equal to zero if $\Delta i_c = 0$. This
means that asymmetry of the shunt resistances itself does not affect
the circulating current, as it is expected for superconducting
state. In this case the nonlinear term of $f(\psi)$ (\ref{f_trans
ib=0}) is multiplied just by $g = \Sigma i_c/2$. Generally, the
effect of difference of the shunt resistances on $g(\psi)$ remains
negligible for arbitrary $\Delta i_c$.

For equal shunt resistances $\Delta r_n = 0$ the $A_I$ constant is
$A_I = -\Delta i_c/\Sigma i_c$ and so the $g(\psi)$ function is the
product of $2i_{c1}i_{c2}/\Sigma i_c$ and $1/\sqrt{1 + (\Delta
i_c/\Sigma i_c)^2\tan^2\psi}$. The first factor is of the order of
unity for $i_{c1}, i_{c2} \approx 1$. The second factor is positive
and less than unity. It has a deep at $\psi = \pi/2 + \pi n$ (where
$n$ is integer). These properties of $g(\psi)$ factors lead to
effective smoothing of nonlinear bend of the $f(\psi)$ function in
the vicinity of $\psi = \pi/2 + \pi n$ (see Fig.~\ref{Fig10}(a)).
The bend smoothing, in turn, leads to shrinking of hysteresis of the
circulating current curve illustrated in Fig.~\ref{Fig10}(b). This
effect increases with increase of $\Delta i_c$. It is manifested in
the manner somewhat analogous to nonzero bias current effect
considered in Section 3. However, here it results from limitation of
the circulating current by smaller critical current of unequal
junctions.

The $f(\psi)$, $f^*(\psi)$ functions shown in Fig.~\ref{Fig10}(a)
are calculated using expressions (\ref{f_trans ib=0}), (\ref{f trans
asym}) for symmetrical case, and for the same parameters of SQUID
asymmetry as the ones taken for calculations of the data presented
in Fig.~\ref{Fig9}(a) correspondingly. The circulating currents
$i_{cir}$, $i^*_{cir}$ in Fig.~\ref{Fig10}(b) are found using $f$,
$f^*$ functions, expressions (\ref{Icir_sup}), (\ref{psi(phie)}),
and derivative of $f^*(\psi)$ function
\begin{multline} \label{f' trans asym}
\frac{\partial f^*}{\partial \psi} = g(\psi)\frac{l}{2}\sgn(\cos\psi)\bigg[\cos\psi\\
- \frac{\sin\psi\tan\psi A_I^2(1 + \tan^2\psi)}{1 +
A_I^2\tan^2\psi}\bigg] + 1.
\end{multline}
Numerical data shown in Fig.~\ref{Fig10}(b) are obtained using
system (\ref{asymSQeq}).

\subsubsection{Resistive state.}

Time-averaged circulating current in resistive state can be found
using expressions (\ref{SolFor0ap*}), (\ref{wJ0*}), (\ref{Psi1*})
and (\ref{Psi0*total}):
\begin{multline} \label{i cir res asym}
\overline{i}^*_{cir} = \frac{\left[\Sigma v_c\Delta v_c - 2
\frac{\Sigma l w^*_{J-}}{\Sigma r_n} v_{c1}v_{c2} \sin 2\psi^*_-
\right]\left(\frac{i^*_b}{2} -  w^*_{J-}\right)}{2 \Sigma r_n
\left(\left[\frac{\Sigma l w^{*}_{J-}}{\Sigma r_n}\right]^2 +
1\right)\left[\frac{\Delta v_c^2}{4} + v_{c1}v_{c2}\cos^2\psi^*_-
\right]}\\ - \frac{i_b}{2}\frac{\Delta r_n}{\Sigma r_n}.
\end{multline}

For symmetrical case this expression converts into (\ref{IcirR1fin})
with corresponding scaling coefficients. However, contrary to
(\ref{IcirR1fin}) here the time-averaged circulating current, in
general, has some offset due to redistribution of the bias current.
For the bias current equal to the SQUID critical current $i_b =
\Sigma i_c$ the averaged circulating current at $\psi^*_- = 0$ is
\begin{equation}\label{i cir res asym psi0}
\overline{i}^*_{cir} = \frac{1}{\Sigma r_n}\left(\Delta v_c -
\frac{\Sigma i_c}{2}\Delta r_n\right).
\end{equation}
For $\Delta i_c = 0$ the current is $\overline{i}^*_{cir} = 0$ as it
is expected, but for $\Delta r_n = 0$ the current is
$\overline{i}^*_{cir} = \Delta i_c /2$.

Note, that in contrast with (\ref{IcirR1fin}) the current (\ref{i
cir res asym}) for $l = 0$ is not zero. At the peak of the voltage
response $\psi^*_- = -\pi/2$ and for $i_b = \Sigma i_c$ expression
(\ref{i cir res asym}) can be simplified if $\Delta i_c$ or $\Delta
r_n$ is zero.

For $\Delta i_c = 0$ it is
\begin{equation}\label{i cir res asym L0 dr}
\overline{i}^*_{cir} = -\frac{\Delta r_n}{2}\frac{\Sigma i_c}{\Sigma
r_n}\left(1 - \frac{\Sigma r_n}{\Sigma r_n + 2\sqrt{r_{n1}r_{n2}}}
\right).
\end{equation}
Remembering that $\Delta v_c = \Sigma i_c \Delta r_n/2$ and $\Sigma
v_c = \Sigma i_c \Sigma r_n/2$, one can see that (\ref{i cir res
asym L0 dr}) is just the second term of the expression
(\ref{VppAsymL0}) for the voltage response amplitude without the
factor $\Delta r_n/2$ arising from the first term of (\ref{wJ1*e}).
Since the expression in the brackets (\ref{i cir res asym L0 dr}) is
always positive, the averaged circulating current in this case
decreases the voltage response because it always flows toward the
junction with smaller shunt resistor. Comparing (\ref{i cir res asym
psi0}) and (\ref{i cir res asym L0 dr}) we conclude that the applied
flux increases the averaged circulating current induced due to
inequality of the shunt resistances.

For $\Delta r_n = 0$ the averaged circulating current is
\begin{equation}\label{i cir res asym L0 dic}
\overline{i}^*_{cir} = \frac{\Delta i_c}{2}\frac{\Sigma i_c}{\Sigma
i_c + 2\sqrt{i_{c1}i_{c2}}}.
\end{equation}
Since the second factor in (\ref{i cir res asym L0 dic}) is less
than unity, it is seen that the applied flux decreases the averaged
circulating current induced due to inequality of the critical
currents (compare (\ref{i cir res asym psi0}) and (\ref{i cir res
asym L0 dic})).

\begin{figure}[t]
\resizebox{1\columnwidth}{!}{
\includegraphics{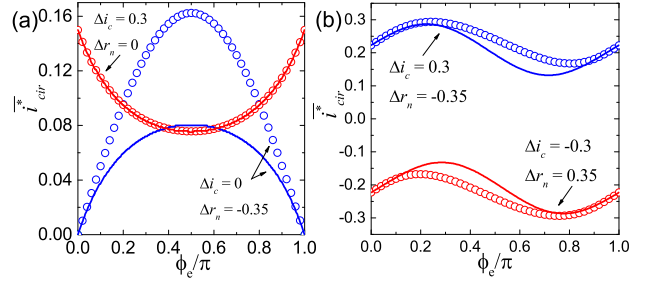}}
\caption{The averaged circulating current $\overline{i}^*_{cir}$
versus the external flux $\phi_e$ calculated using expression
(\ref{i cir res asym}) (solid lines) and obtained using numerical
calculations of system (\ref{asymSQeq}) (dots). (a) $\Delta i_c =
0$, $\Delta r_n = -0.35$ and $\Delta i_c = 0.3$, $\Delta r_n = 0$ at
$i_b = \Sigma i_c = 1.9$, $\Sigma r_n = 2.05$, $l = 0$. (b) $\Delta
i_c = -0.3$, $\Delta r_n = 0.35$ and $\Delta i_c = 0.3$, $\Delta r_n
= -0.35$ at $i_b = 2.45$, $\Sigma i_c = 1.9$, $\Sigma r_n = 2.05$,
$\Sigma l = 1$ and $\Delta l = 0$.} \label{Fig11}
\end{figure}

Figure~\ref{Fig11}(a) presents the averaged circulating current
(\ref{i cir res asym}) calculated for $\Delta i_c = 0$, $\Delta r_n
= -0.35$ and $\Delta i_c = 0.3$, $\Delta r_n = 0$ at $i_b = \Sigma
i_c = 1.9$, $\Sigma r_n = 2.05$, $l = 0$ (solid lines).
Corresponding data calculated numerically using system
(\ref{asymSQeq}) are shown by dots. It is seen that even for small
absolute current values $\overline{i}^*_{cir}$ the data
corresponding to the averaged circulating current in SQUID with
asymmetry of the critical currents are perfectly consistent.
Although the data for SQUID with asymmetry of the shunt resistances
differ noticeably at $\phi_e \approx \pi/2$ resulting from
approximate solution for $\psi^*_\sim$ (\ref{Psi1*}), qualitative
behavior is still found correctly for both asymmetries.

Figure~\ref{Fig11}(b) shows the circulating current calculated for
the same set of parameters as the ones taken for Fig.~\ref{Fig9}(a)
but $i_b = 2.45$ ($\Delta l = 0$), and for SQUID with inverted
asymmetry of the critical currents and shunt resistances ($\Delta
i_c \rightarrow -\Delta i_c$, $\Delta r_n \rightarrow -\Delta r_n$,
while $\Sigma i_c$, $\Sigma r_n$ are held the same). The data
obtained by expression (\ref{i cir res asym}) are shown by lines,
and the ones calculated numerically using (\ref{asymSQeq}) are
presented by dots. While consistency of the curves remains to be
just qualitative, the offset to the averaged circulating current is
found precisely.

\section{SQUID with inductance of practical device}

The voltage-flux and current-flux functions obtained in Sections 2,
3 for resistive state are valid only for SQUID with small inductance
$l \leq 1$. In this Section we fit numerical data for the SQUID
responses by our analytical expressions introducing fitting
parameters to expand the frame of validity of our approach to higher
values of the inductance. This makes our expressions suitable for
design process of practical SQUIDs and SQUID-based circuits.

\subsection{SQUID voltage response}

\begin{figure}[t]
\begin{center}
\includegraphics[width=6cm]{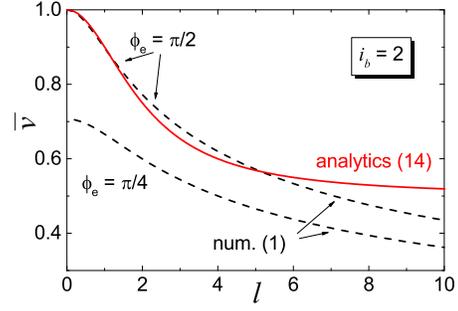}
\end{center}
\caption{Time-averaged voltage on symmetrical SQUID at $i_b = 2$
calculated using (\ref{VppSym}) at $\phi_e = \pi/2$ (solid line) and
obtained by numerical calculations of (\ref{SQUIDeq}) at $\phi_e =
\pi/2,~ \pi/4$ (dotted lines).} \label{Fig12}
\end{figure}

In Section 2 it was considered that expression (\ref{VppSym})
obtained for the voltage response amplitude is relevant only for
small values of the inductance $l \leq 1$. Since the voltage
$\overline{v}$ across symmetrical SQUID at $i_b = 2$ and $\phi_e =
0$ is zero, the response amplitude is just $\overline{v}$ at $\phi_e
= \pi/2$. Comparison of dependencies $\overline{v}(l)$ calculated
using (\ref{VppSym}) and obtained by numerical calculations of
system (\ref{SQUIDeq}) is shown in Fig.~\ref{Fig12}. It is seen that
proposed linearization of (\ref{SQUIDeq}) leads to inaccurate
determination of the voltage response amplitude for $l > 1$ that
limits application of the derived expressions. To make our approach
usefull in practical circuits design we propose usage of expression
(\ref{SQwjtot1}) for SQUID voltage response with fitting parameters
which can be found from numerical calculations of (\ref{SQUIDeq}) as
follows.

Dependence of the voltage on the inductance $\overline{v}(l)$ at
$i_b = 2$, $\phi_e = \pi/2$ obtained by numerical calculations of
(\ref{SQUIDeq}) and presented in Fig.~\ref{Fig12} can be well
approximated by function
\begin{equation}\label{vpp num}
\overline{v}_{\pi/2}^{num} = 1 - p,
\end{equation}
where
\begin{equation}\label{p}
p = \frac{l^{1.66}}{2.44 l^{1.48} + 7.22}.
\end{equation}

The voltages (\ref{VppSym}) and (\ref{vpp num})  can be equalized
with introduction of effective inductance $l_a$:
\begin{equation*}
1 - \frac{l_a^2}{2 l_a^2 + 8} = 1 - p,
\end{equation*}
which gives
\begin{equation}\label{la}
l_a = 2\sqrt{\frac{2p}{1 - 2p}}.
\end{equation}
This inductance being substituted in (\ref{SQwjtot1}) instead of $l$
allows calculation of the voltage response with correct amplitude.
Unfortunately, the shape of the voltage response is still not
reproduced at this stage.

Improving of consistency of the shape is possible by allowing two
fitting parameters, the effective inductance $l_s$ and the amplitude
$A$ for the second term of (\ref{SQwjtot1}):
\begin{equation}\label{SQwjtotFit}
\overline{v}(\phi_e) = w_{J-} - A\frac{l_s^2 w^2_{J-}}{l_s^2
w^2_{J-}+4} \left(\frac{i_b}{2} - w_{J-}\right)\tan^2\phi_e.
\end{equation}

To define both parameters we introduce another fitting dependence,
namely $\overline{v}(l)$ at $i_b = 2$, $\phi_e = \pi/4$ calculated
numerically, which is also shown in Fig.~\ref{Fig12}. This
dependence can be approximated by function
\begin{equation}\label{v num piO4}
\overline{v}^{num}_{\pi/4} = \frac{\sqrt{2}}{2} - q,
\end{equation}
where
\begin{equation}\label{q}
q = \frac{l^{1.92}}{5.25 l^{1.625} + 19.14}.
\end{equation}
Both fitting parameters $l_s$, $A$ can be derived from system
obtained by equating (\ref{SQwjtotFit}) and (\ref{vpp num}), (\ref{v
num piO4}) at $\phi_e = \pi/2,~\pi/4$ correspondingly.

We note that amplitude of the voltage response is found well enough
with $l_a$ even for $i_b \neq 2$ that allows introduction of
dependence of desired parameters on the bias current. The equation
for $\phi_e = \pi/2$ can accordingly be written for arbitrary $i_b$
with the found $l_a$, and so the system for definition of the
desired parameters reduces to
\begin{subequations} \label{A ls sys}
\begin{equation}
A \frac{l_s^2 i_b}{l_s^2 i_b^2 + 16} = \frac{l_a^2 i_b}{l_a^2 i_b^2
+ 16},
\end{equation}
\begin{equation}
A \frac{l_s^2}{l_s^2 + 8}\left(1 - \frac{\sqrt{2}}{2}\right) = q.
\end{equation}
\end{subequations}
Solution of the system (\ref{A ls sys}) for $l_s$, $A$ gives
\begin{subequations} \label{A ls}
\begin{equation}
l_s = 4\sqrt{\frac{2(q-p) + pq(i_b^2 - 4) + \sqrt{2}p}{2(i_b^2 p -
2q) - 2pq(i_b^2 - 4) - i_b^2 \sqrt{2} p}},
\end{equation}
\begin{equation}
A = \frac{p q (i_b^2 - 2)}{2(q - p - 2pq) + p(\sqrt{2} + i_b^2 q)}.
\end{equation}
\end{subequations}
Frame of validity of this definition corresponds to the range where
expression under the square root in (\ref{A ls}a) is positive. This
range is $l \in [0.35, 6.85]$ for $i_b = 2$ and it increases with
increase of the bias current.
\begin{figure}[t]
\centering
\begin{minipage}[t]{0.49\columnwidth}
\resizebox{1\columnwidth}{!}{
\includegraphics{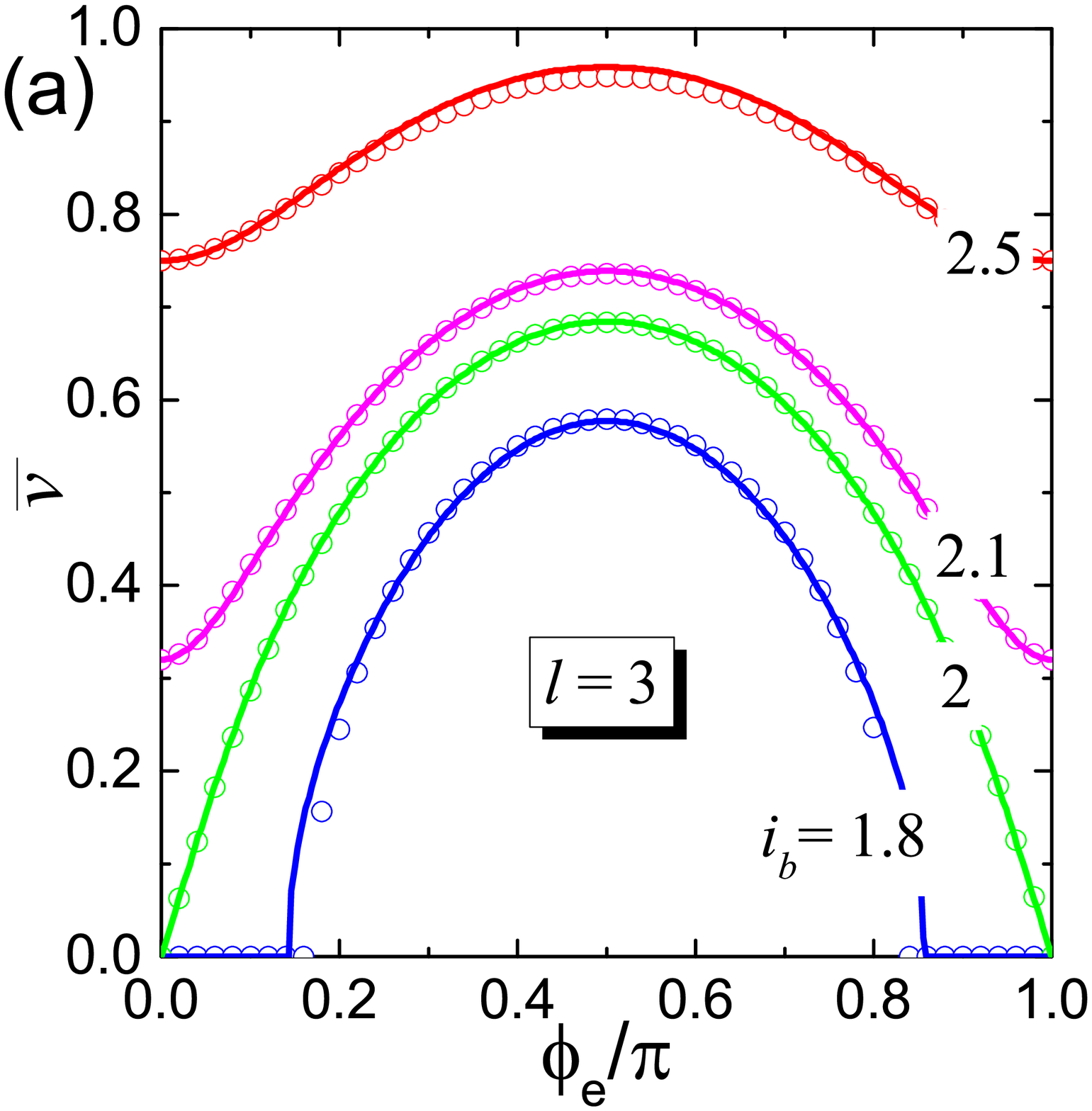}}
\end{minipage}
\begin{minipage}[t]{0.49\columnwidth}
\resizebox{1\columnwidth}{!}{
\includegraphics{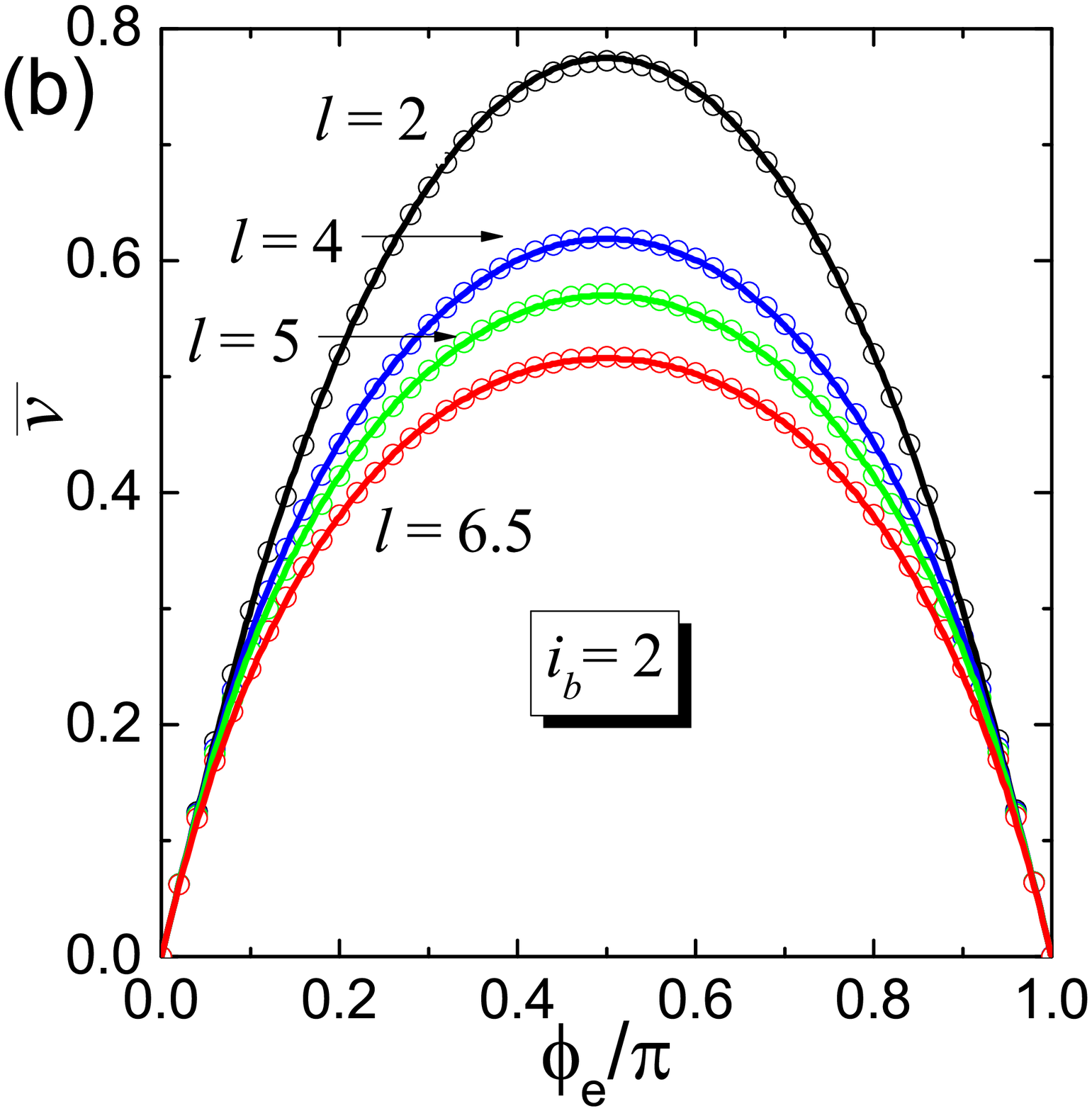}}
\end{minipage}
\begin{minipage}[t]{0.49\columnwidth}
\resizebox{1\columnwidth}{!}{
\includegraphics{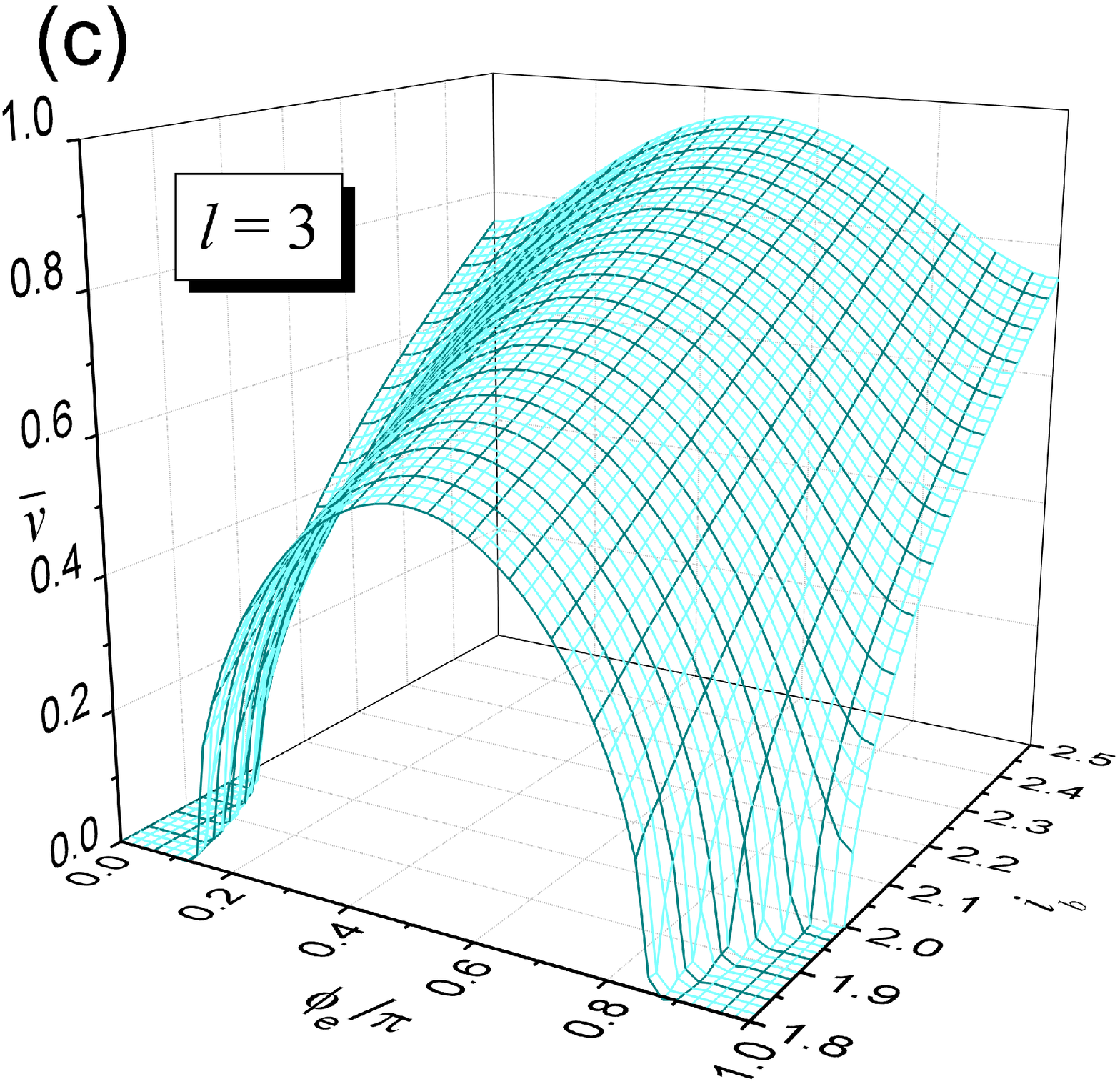}}
\end{minipage}
\begin{minipage}[t]{0.49\columnwidth}
\resizebox{1\columnwidth}{!}{
\includegraphics{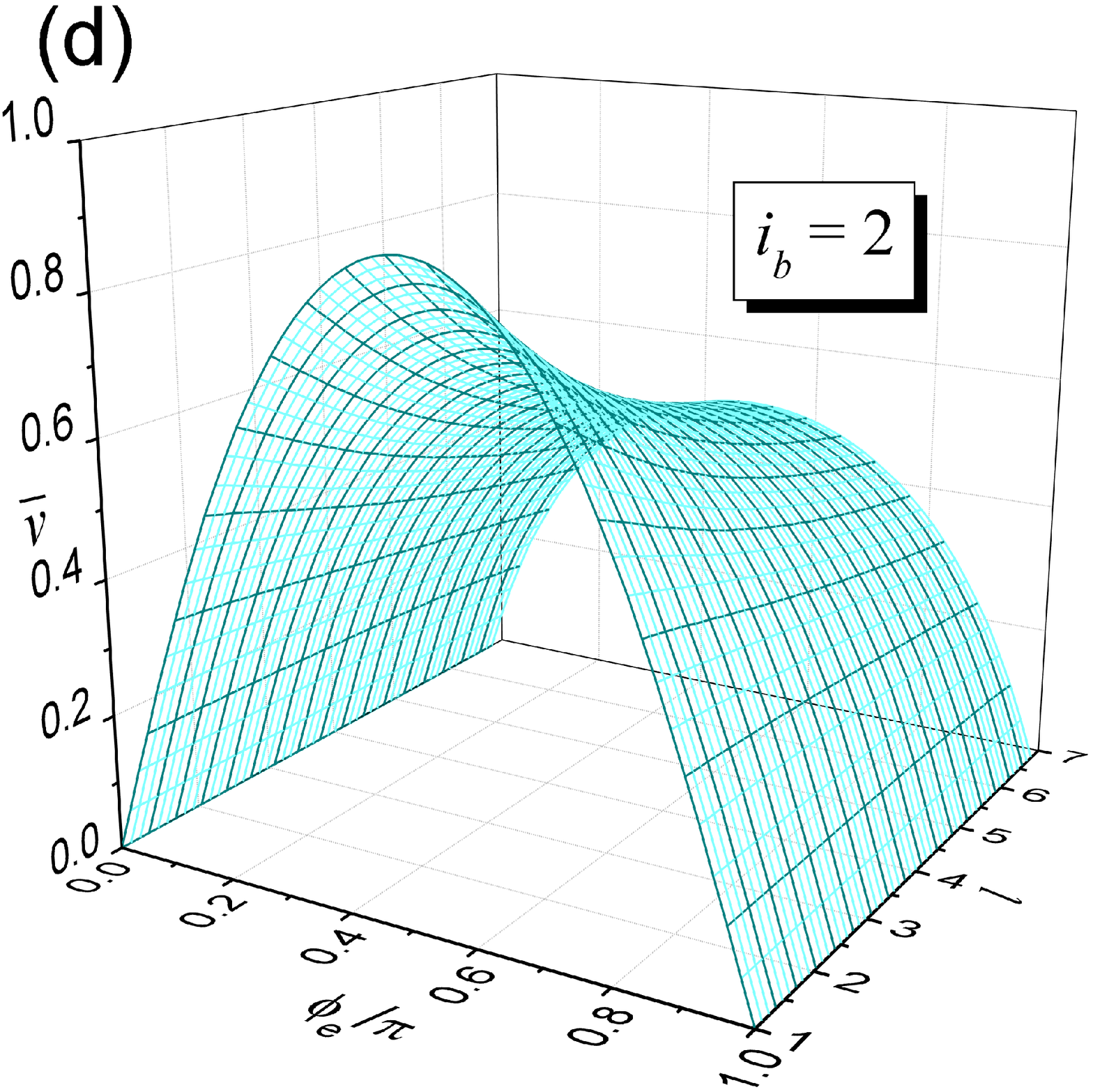}}
\end{minipage}
\caption{Comparison of SQUID voltage response $\overline{v}(\phi_e)$
obtained using (\ref{SQwjtotFit}) (lines) and using numerical
calculations of system (\ref{SQUIDeq}) (dots) at practical
inductance value $l = 3$ for a set of the bias currents $i_b = 1.8,
2, 2.1, 2.5$ (a), and at $i_b = 2$ for a set of the inductance
values $l = 2, 4, 5, 6.5$ (b). SQUID voltage response versus $i_b$
at $l = 3$ (c), and versus $l$ at $i_b = 2$ (d) calculated by
(\ref{SQwjtotFit}).} \label{Fig13}
\end{figure}

Figure~\ref{Fig13}(a),(b) present comparison of the data obtained by
expression (\ref{SQwjtotFit}) (lines) and using numerical
calculations of system (\ref{SQUIDeq}) (dots). It is seen that the
data are well consistent especially in the vicinity of the SQUID
critical current $i_b \approx 2$. This result allows one to study
the voltage response (or IV-curve) as a function of the parameters
$i_b$, $\phi_e$, $l$, and use the data for practical purpose.
Corresponding examples of 3D plots are shown in
Fig.~\ref{Fig13}(c),(d).

\subsection{SQUID current response}

Time-averaged circulating current in symmetrical SQUID can be found
similarly using numerical calculations of system (\ref{SQUIDeq}) and
expression (\ref{IcirR1fin}). Obtained expression for the current
response can be rewritten as
\begin{equation}\label{SQicirtotFit}
\overline{i}_{cir}(\phi_e) = A_{i}\frac{2 l_{si} w_{J-}}{l_{si}^2
w_{J-}^2 + 4}\left(\frac{i_b}{2} - w_{J-}\right)\tan\phi_e,
\end{equation}
where $l_{si}$, $A_i$ are the fitting parameters. While the
parameters are defined for arbitrary inductances and bias currents,
their expressions are rather complicated
\begin{equation}\label{Ai}
A_i = \frac{(\frac{i_b}{2})^\beta
l^{0.87}}{([1-\alpha](\frac{i_b}{2})^\gamma + \alpha)(0.91 l^{1.4} +
2.26)},
\end{equation}
where
\begin{equation*}
\alpha = \frac{l^{2.32}}{1.4 l^{2.39} + 0.31},
\end{equation*}
\begin{equation*}
\beta,\gamma =
\frac{\sqrt{2}-1}{f_{\beta,\gamma}}\left(1+\sqrt{1-\frac{32
f^2_{\beta,\gamma}}{12 - 8\sqrt{2}}}\right),
\end{equation*}
\begin{equation*}
f_{\beta} = \frac{l^{2.97} + 0.69}{6.84 l^{3.35} +
6.53},~~f_{\gamma} = \frac{l^{2.84} + 1.15}{6.68 l^{3.22} + 9.21},
\end{equation*}
and
\begin{equation}\label{lsi}
l_{si} = \frac{\sqrt{2}-1}{f_l}\left(1+\sgn(l - 1.3)\sqrt{1-\frac{32
f^2_l}{12 - 8\sqrt{2}}}\right),
\end{equation}
\begin{equation*}
f_l = \frac{l^{2.03} + 1.2}{5.42 l^{2.34} + 9.81}.
\end{equation*}
\begin{figure}[t]
\centering
\begin{minipage}[t]{0.49\columnwidth}
\resizebox{1\columnwidth}{!}{
\includegraphics{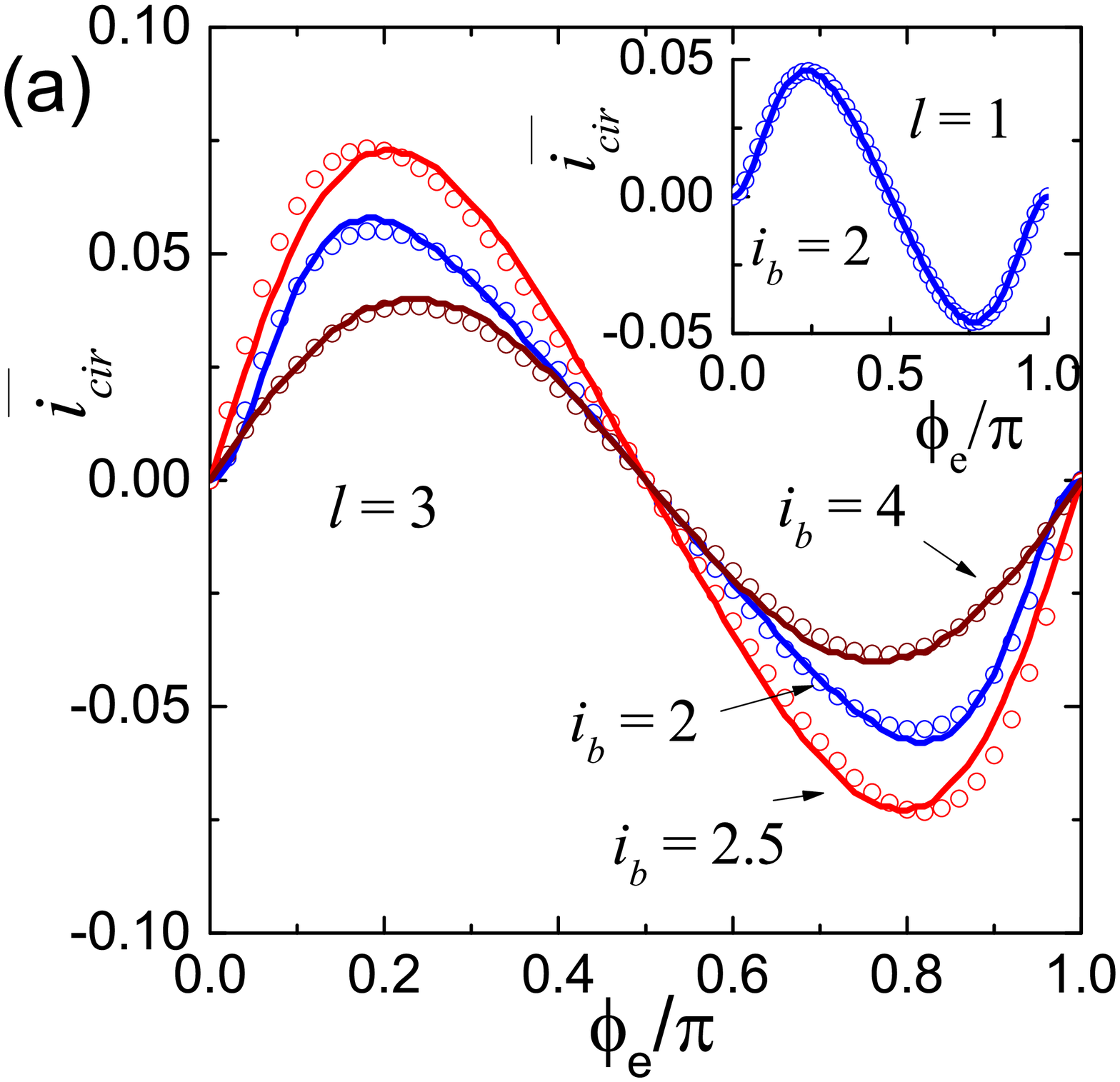}}
\end{minipage}
\begin{minipage}[t]{0.49\columnwidth}
\resizebox{1\columnwidth}{!}{
\includegraphics{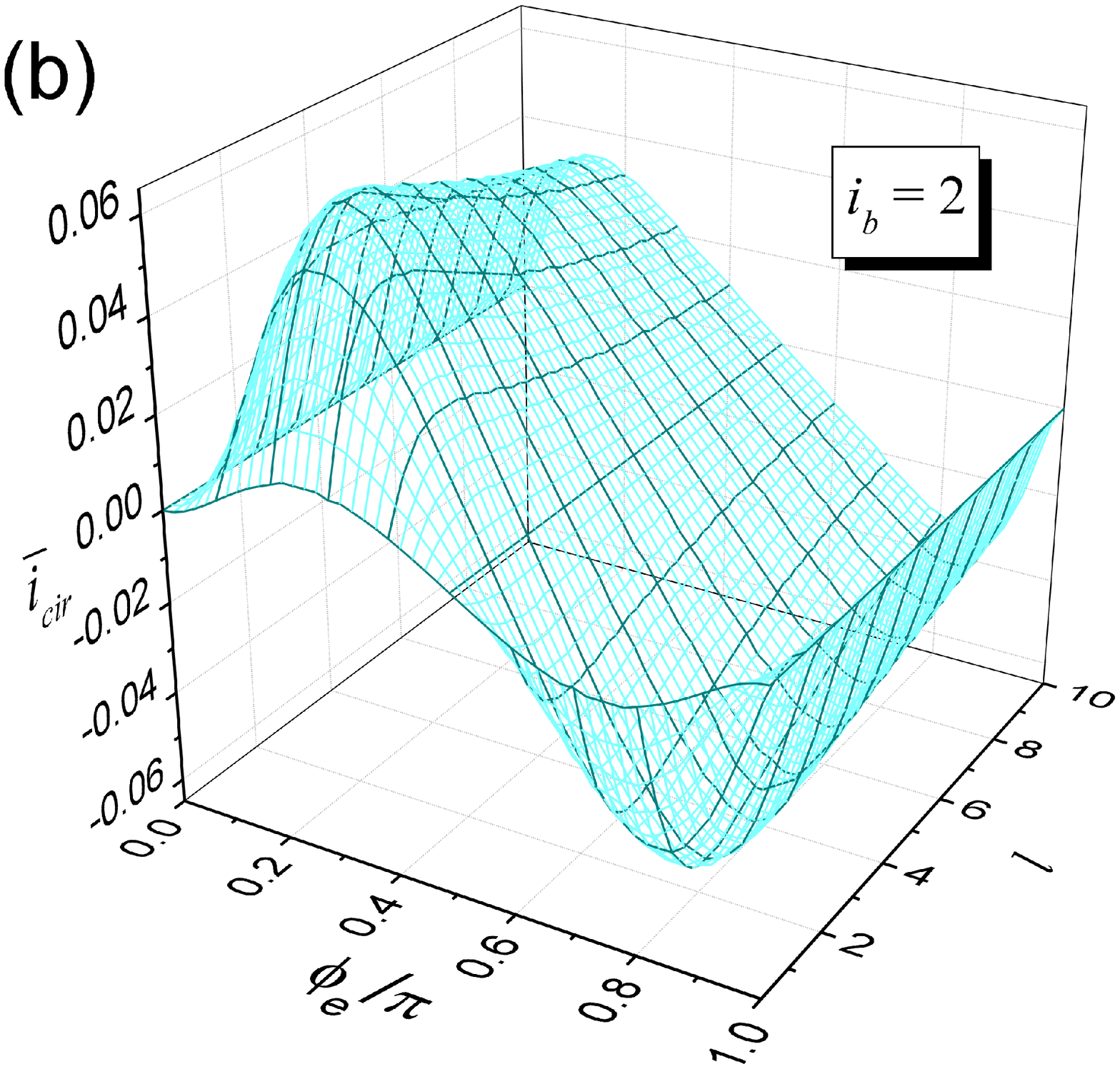}}
\end{minipage}
\begin{minipage}[t]{0.49\columnwidth}
\resizebox{1\columnwidth}{!}{
\includegraphics{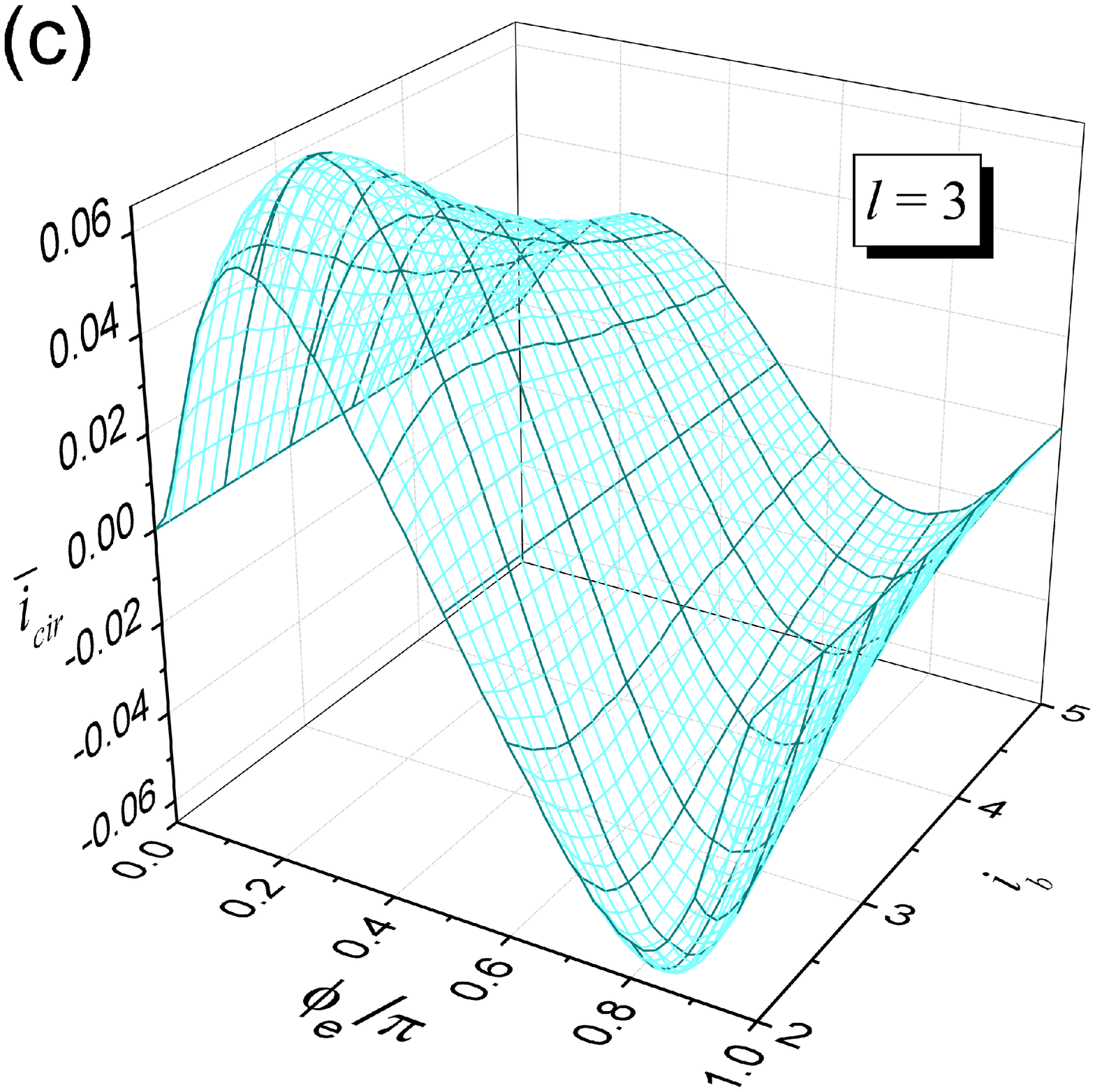}}
\end{minipage}
\begin{minipage}[t]{0.49\columnwidth}
\resizebox{1\columnwidth}{!}{
\includegraphics{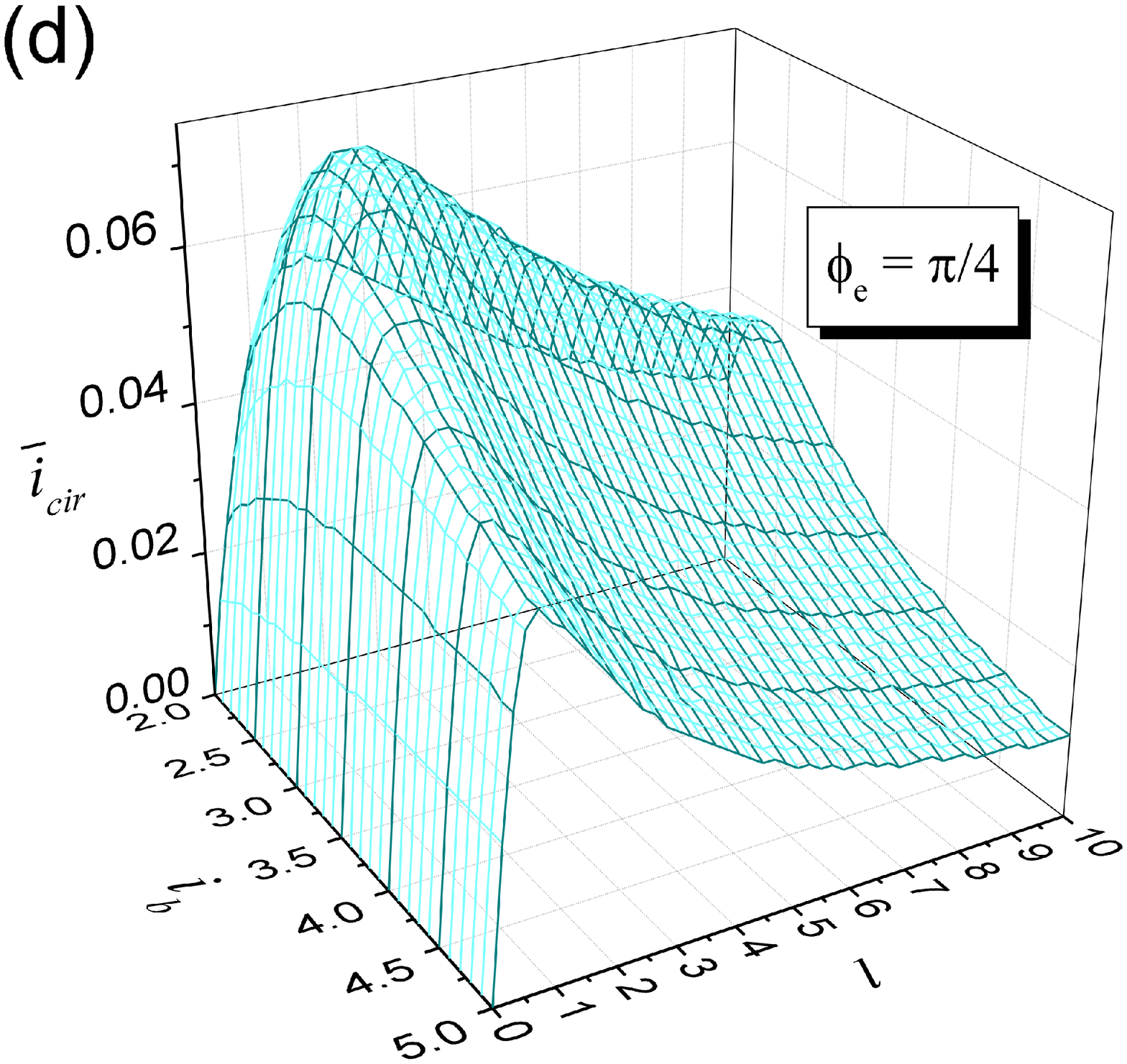}}
\end{minipage}
\caption{(a) Comparison of SQUID current response
$\overline{i}_{cir}(\phi_e)$ obtained by (\ref{SQicirtotFit})
(lines) and using numerical calculations of system (\ref{SQUIDeq})
(dots) at the inductance value $l = 3$ for a set of the bias
currents $i_b = 2, 2.5, 4$. Inset shows curves for $l = 1$, $i_b =
2$. SQUID current response versus $l$ at $i_b = 2$ (b), and versus
$i_b$ at $l = 3$ (c) calculated by (\ref{SQicirtotFit}). (d) The
averaged circulating current $\overline{i}_{cir}$
(\ref{SQicirtotFit}) at $\phi_e = \pi/4$ versus the bias current and
the inductance.} \label{Fig14}
\end{figure}

Figure~\ref{Fig14}(a) shows the time-averaged circulating current
calculated at different bias currents and inductances using
(\ref{SQicirtotFit}) in comparison with corresponding numerical
calculations of system (\ref{SQUIDeq}). The results indicate that
the numerical data are fitted by (\ref{SQicirtotFit}) fairly well.

Dependencies of the SQUID current response versus the inductance and
versus the bias current calculated by (\ref{SQicirtotFit}) are
presented in Fig.~\ref{Fig14}(b),(c). It is interesting to note that
the time-averaged circulating current has maximum over parameters
$\phi_e$, $l$, $i_b$. Nonzero value of $\overline{i}_{cir}$ means
that integral over time of the current circulating in the SQUID in
the time interval between switchings of the junctions (when a fluxon
is passing by the SQUID) is greater than the one corresponding to
leveling of the currents flowing through the junctions which happens
after their switchings.

Obviously, there is no circulating current if the junctions switch
simultaneously ($\phi_e = 0$), while if they switch in antiphase
($\phi_e = \pi/2$) the corresponding integrals are equalized. This
gives optimum value for $\phi_e$ maximizing $|\overline{i}_{cir}|$
about $\phi_e = \pi/4$.

The current $\overline{i}_{cir}$ at this applied flux versus both
remaining parameters $l$, $i_b$ are shown in Fig.~\ref{Fig14}(d).
Optimum value for the inductance outflows from the fact that for $l
\rightarrow 0$ the junctions are synchronized inphase, and the
circulating current is purely symmetric over oscillation period. At
the same time, for $l \rightarrow \infty$ the circulating current is
negligible due to $LR$-filtering.

The bias current is taken into consideration through the oscillation
frequency $w_J$. For small bias current the frequency is low, and so
the time left after switchings for leveling of the currents flowing
through the junctions prevail over the time of fluxon passage, that
equalizes impacts of the integrals. High bias currents mean high
frequencies $w_J$ at which the circulating current is suppressed by
$LR$-filtering as it was considered in Section 2.

Certain values of the parameters providing maximum time-averaged
circulating current can be found from (\ref{SQicirtotFit}) as
\{$\phi_e$, $l$, $i_b$\} = \{$0.21/\pi$, $2.1$, $2.56$\}, at which
the current is $\overline{i}_{cir} = 0.076$. The obtained result
shows that this current can be safely omitted in symmetrical case in
consideration of more complex circuits e.g. a bi-SQUID as it was
done in the work \cite{KSSM}.

\section{Discussion}

Results presented in this paper provide qualitative and quantitative
understanding of processes involved in formation of DC SQUID voltage
and current responses. Obtained expressions of the voltage-flux and
current-flux functions for practical values of SQUID inductance can
be used in design of SQUID-based circuits.

One of difficult problem is modeling of large SQUID arrays. An array
containing $2400$ Josephson junctions was utilized in recent work
\cite{AETBH} as electrically small antenna capable of detecting
radio frequencies from distant sources. While shape of the voltage
response is one of the most important characteristics of such
structure its numerical calculation is quite time consuming. Indeed
standard tools for superconductor circuit simulations as PSCAN
practically limit circuit complexity to $\sim 500$ junctions, so
complex schematics have to be broken into several sub-schematics
\cite{MRep}. In this respect the proposed analytical approach to
describe SQUID time-averaged characteristics provides valuable
solution.

\begin{figure}[t]
\resizebox{1\columnwidth}{!}{
\includegraphics[]{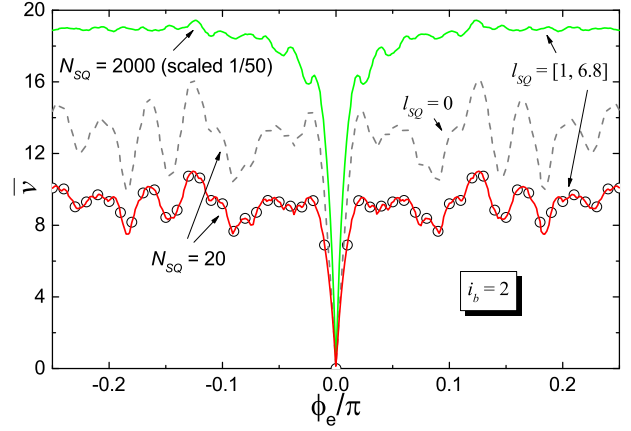}}
\caption{Voltage responses of SQIF structures with normal
distribution of inductances of their cells in the range $l_{SQ} \in
[1,6.8]$. Area of the cells are given as $a_{SQ} = l^2_{SQ}$. Curves
obtained using presented analytical approach (expression
(\ref{SQwjtotFit})) for SQIF structures with number of cells $N_{SQ}
= 20,~2000$ are shown by solid lines (the last curve is scaled
$1/50$). Dots show data for the SQIF with $N_{SQ} = 20$ calculated
numerically using PSCAN. Analytical curve calculated in zero
inductance approximation ($l_{SQ} = 0$) for this SQIF with $N_{SQ} =
20$ is presented by dotted line.} \label{Fig15}
\end{figure}

To illustrate the applicability of our approach to design SQUID
arrays we calculate the voltage responses of serial SQIFs with the
uniform distribution of inductances of their cells in the range
$l_{SQ} \in [1,6.8]$ at the bias current equal to SQIF's critical
current $i_b = 2$, see Fig.~\ref{Fig15}. It is assumed that area of
a cell is proportional to square of it's inductance, and so magnetic
flux applied to each cell is multiplied by $l_{SQ}^2$ in our
calculations. Curve calculated numerically in PSCAN software tool
for SQIF with $N_{SQ} = 20$ SQUID cells is presented by dots. SQIF
responses obtained as a sum of cell responses using
(\ref{SQwjtotFit}) for the same structure and for SQIF with $N_{SQ}
= 2000$ cells are shown by solid lines (the last curve is scaled
down $50$ times). To display effect of the inductances we calculate
the response of the SQIF structure with $N_{SQ} = 20$ putting
$l_{SQ} = 0$ and keeping areas of the cells unchanged (dashed line).

It is seen that data calculated numerically and calculated using our
analytical approach are perfectly consistent. At the same time, the
curve calculated in zero inductance approximation (for $l_{SQ} = 0$)
can provide only rough estimation of the voltage response amplitude
and shape. We got no significant delay in time of calculation of the
curve for the SQIF structure with $N_{SQ} = 2000$ cells compared to
the ones with $N_{SQ} = 20$ using our approach. It took about a
second on conventional laptop, while corresponding time of numerical
calculation would be at least three orders greater. This means that
our analytical expressions can be readily used for optimization of
such complex circuits.

\section{Conclusion}

In conclusion, we have developed analytical approach for calculation
of DC SQUID voltage and current responses in resistive state for
inductance in the range $l \leq 1$. Using two fitting parameters we
expanded the frame of validity of our approach to practical values
of the inductance up to $l \approx 7$. The circulating current in
superconducting state was found for arbitrary values of the
inductance. We considered effect provided by technological spread of
SQUID parameters relevant for LTS technology generalizing our
approach to a case of slightly different critical currents and shunt
resistances of SQUID junctions, and unequal SQUID inductive
shoulders. We showed that our analytical expressions can be used for
calculation of practical SQUID and serial SQUID array responses,
confirming this by comparison with numerical calculation results.

\ack Authors are grateful to Matthias Schmelz for careful reading of
the paper and valuable suggestions. This work was supported by
President of Russian Federation grant MK-5813.2016.2, RFBR grants
No. 15-32-20362-mol$\_$a$\_$ved, 16-29-09515-ofi$\_$m, Ministry of
Education and Science of the Russian Federation, grant No.
14.Y26.31.0007, and grant for leading scientific school No.
8168.2016.2. Mikhail Kupriyanov would like to thank support of
Program of Competitive Growth at Kazan Federal University.

Development of analytical description of symmetrical SQUID voltage
and current responses was performed under support of RFBR grant No.
15-32-20362-mol$\_$a$\_$ved. Study of effect of small technological
spread of parameters on SQUID responses was done in the frame of
President of Russian Federation grant MK-5813.2016.2. Applicability
of developed analytical expressions to practical LTS SQUID-based
circuits design was studied under support of RFBR grant No.
16-29-09515-ofi$\_$m.


\section*{References}
\begin{thebibliography}{99}

\bibitem{CB} Clarke J, Braginsky A I 2004 {\it The SQUID Handbook} (Weinheim: Wiley) Vol. 1
\bibitem{W} Weinstock H (ed) 1996 {\it SQUID sensors: Fundamentals, Fabrication and Applications} (Dordrecht: Kluwer) pp 1 –- 62
\bibitem{GV} Granata C, Vettoliere A 2016 Nano Superconducting Quantum Interference device: a powerful tool for nanoscale investigations {\it Phys. Rep.} {\bf 614} 1 -- 69

\bibitem{SZSSALMM} Sch\"onau T, Zakosarenko V, Schmelz M, Stolz R, Anders S, Linzen S, Meyer M, Meyer H-G 2015 A three-axis SQUID-based absolute vector magnetometer {\it Rev. Sci. Instrum.} {\bf 86} 105002
\bibitem{CSSZMM} Chwala A, Stolz R, Schmelz M, Zakosarenko V, Meyer M, Meyer H-G 2015 SQUID Systems for Geophysical Time Domain Electromagnetics (TEM) at IPHT Jena {\it IEICE Trans. Electron.} {\bf E98C} 167 -- 173
\bibitem{SSZSMAFM} Sch\"onau T, Schmelz M, Zakosarenko V, Stolz R, Meyer M, Anders S, Fritzsch L, Meyer H-G 2013 SQUID-based setup for the absolute measurement of the Earth's magnetic field {\it Supercond. Sci. Technol.} {\bf 26} 035013
\bibitem{SSZSAFMMM} Schmelz M, Stolz R, Zakosarenko V, Sch\"onau T, Anders S, Fritzsch L, M\"uck M, and Meyer H-G 2011 Field-stable SQUID magnetometer with sub-fT Hz$^{-1/2}$ resolution based on sub-micrometer cross-type Josephson tunnel junctions {\it Supercond. Sci. Technol.} {\bf 24} 065009
\bibitem{SSZAFRM} Schmelz M, Matsui Y, Stolz R, Zakosarenko V, Sch\"onau T, Anders S, Linzen S, Itozaki H and Meyer H-G 2015 Investigation of all niobium nano-SQUIDs based on sub-micrometer cross-type Josephson junctions {\it Supercond. Sci. Technol.} {\bf 28} 015004

\bibitem{ProM} Prokopenko G V, Mukhanov O A 2013 Wideband microwave low noise amplifiers based on biSQUID SQIFs {\it International Superconductive Electronics Conference (ISEC), 2013 IEEE 14th ISEC, July 7 - 11}
\bibitem{SSZSAFM} Sch\"onau T, Schmelz M, Zakosarenko V, Stolz R, Anders S, Fritzsch L,  Meyer H-G 2012 SQIF-based dc SQUID amplifier with intrinsic negative feedback {\it Supercond. Sci. Technol.} {\bf 25} 015005


\bibitem{KSKMSQIFDriv} Kornev V K, Soloviev I I, Klenov N V, Mukhanov O A 2007 Development of sqif-based output broad band amplifier {\it IEEE Trans. Appl. Supercond.} {\bf 17}(2) 569 -- 572

\bibitem{ZSKHAPMKSESM} Zakosarenko V, Schulz M, Krueger A, Heinz E, Anders S, Peiselt K, May T, Kreysa E, Siringo G, Esch W, Starkloff M, Meyer H-G 2011 Time-domain multiplexed SQUID readout of a bolometer camera for APEX {\it Supercond. Sci. Technol.} {\bf 24} 015011

\bibitem{KSSKM} Kornev V K, Soloviev I I, Sharafiev A V, Klenov N V and Mukhanov O A 2013 Active electrically small antenna based on superconducting quantum array {\it IEEE Trans. Appl. Supercond.} {\bf 23} 1800405
\bibitem{AETBH} de Andrade M C, de Escobar A L, Taylor B J, Berggren S, Higa B, Son Dinh, Fagaly R L, Talvacchio J, Nechay B, Przybysz J 2015 Detection of Far-Field Radio-Frequency Signals by Niobium Superconducting Quantum Interference Device Arrays {\it IEEE Trans. Appl. Supercond.} {\bf 25} 1603005

\bibitem{KSSKSM} Kornev V K, Sharafiev A V, Soloviev I I, Kolotinskiy N V, Scripka V A, Mukhanov O A 2014 Superconducting Quantum Arrays {\it IEEE Trans. Appl. Supercond.} {\bf 24} 1800606

\bibitem{OHS} Oppenl\"ander J, H\"aussler C, and Schopohl N, 2000 Non-Phi(0)-periodic macroscopic quantum interference in one - dimensional parallel Josephson junction arrays with unconventional grating structure {\it Phys. Rev. B} {\bf 63} 024511
\bibitem{HOS} H\"aussler C, Oppenl\"ander J, and Schopohl N, 2001 Nonperiodic flux to voltage conversion of series arrays of dc superconducting quantum interference devices {\it J. Appl. Phys.} {\bf 89} 1875

\bibitem{KSKM09} Kornev V K, Soloviev I I, Klenov N V and Mukhanov O A 2009 Bi-SQUID— novel linearization method for dc SQUID voltage response {\it Supercond. Sci. Technol.} {\bf 22} 114011

\bibitem{KSKM11} Kornev V K, Soloviev I I, Klenov N V and Mukhanov O A 2011 Design and Experimental Evaluation of SQIF Arrays With Linear Voltage Response {\it IEEE Trans. Appl. Supercond.} {\bf 21} 394 -- 398

\bibitem{MKVFK} Mukhanov O A, Kirichenko D, Vernik I V, Filippov T V, Kirichenko A, Webber R, Dotsenko V, Talalaevskii A, Tang J C, Sahu A, Shevchenko P, Miller R, Kaplan S B, Sarwana A and Gupta D 2008 Superconductor Digital-RF Receiver Systems {\it IEICE Trans. Electron.} {\bf E91–C} 306 -- 317

\bibitem{KSKM10} Kornev V K, Soloviev I I, Klenov N V and Mukhanov O A 2010 Progress in high-linearity multi-element Josephson structures {\it Physica C} {\bf 470} 886 -- 889
\bibitem{KSKSM} Kornev V K, Soloviev I I, Klenov N V, Sharafiev A V and Mukhanov O A 2011 Linear bi-SQUID arrays for electrically small antennas {\it IEEE Trans. Appl. Supercond.} {\bf 21} 713 -- 716
\bibitem{LBEPR11} Longhini P, Berggren A, Palacios A, In V, de Escobar A L 2011 Modeling Non-Locally Coupled DC SQUID Arrays {\it IEEE Trans. Appl. Supercond.} {\bf 21} 391 -- 393
\bibitem{SSKSS} Sharafiev A, Soloviev I, Kornev V, Schmelz M, Stolz R, Zakosarenko V, Anders S and Meyer H-G 2012 Bi-SQUIDs with submicron cross-type Josephson tunnel junctions {\it Supercond. Sci. Technol.} {\bf 25} 045001
\bibitem{LBEPR} Longhini P, Berggren A, de Escobar A L, Palacios A, Rice S, Taylor B, In V, Mukhanov O A, Prokopenko G, Nisenoff M, Wong E and De Andrade M C 2012 Voltage response of non-uniform arrays of bi-superconductive quantum interference devices {\it Journ. Appl. Phys.} {\bf 111} 093920
\bibitem{PMETA} Prokopenko G V, Mukhanov O A, de Escobar A L, Taylor B, de Andrade M C, Berggren S, Longhini P, Palacios A, Nisenoff M and Fagaly R L 2013 DC and RF measurements of serial bi-SQUID arrays {\it IEEE Trans. Appl. Supercond.} {\bf 23} 1400607
\bibitem{BPLPM} Berggren S, Prokopenko G, Longhini P, Palacios A, Mukhanov O A, de Escobar A L, Taylor B J, de Andrade M C, Nisenoff M, Fagaly R L, Wong T, Cho E, Wong E, In V 2013 Development of 2D bi-SQUID arrays with high linearity {\it IEEE Trans. Appl. Supercond.} {\bf 23} 1400208
\bibitem{WCAD} Wu S M, Cybart S A, Anton S M, Dynes R C 2013 Simulation of Series Arrays of Superconducting Quantum Interference Devices {\it IEEE Trans. Appl. Supercond.} {\bf 23} 1600104
\bibitem{KSSM} Kornev V K, Sharafiev A V, Soloviev I I and Mukhanov O A 2014 Signal and noise characteristics of bi-SQUID {\it Supercond. Sci. Technol.} {\bf 27} 115009

\bibitem{BP} Barone A, Paterno G 1982 {\it Physics and applications of the Josephson effect} (New York: Wiley)
\bibitem{L} Likharev K K 1986 {\it Dynamics of Josephson junctions and circtuis} (Amsterdam: Gordon and Breach)

\bibitem{C98} Chesca B 1998 Analytical theory of DC SQUIDS operating in the presence of thermal fluctuations {\it J. Low. Temp. Phys.} {\bf 112} 165 -- 196
\bibitem{C99} Chesca B 1999 The effect of thermal noise on the I–V curves of high inductance dc SQUIDs in the presence of microwave radiation {\it J. Low. Temp. Phys.} {\bf 116} 167 -- 186
\bibitem{C991} Chesca B 1999 The effect of thermal fluctuations on the operation of DC SQUIDs at 77 K – a fundamental analytical approach {\it IEEE Trans. Appl. Supercond.} {\bf 9} 2955 -- 2960
\bibitem{G02} Greenberg Ya S 2002 Theory of the voltage–current characteristic of high Tc DC SQUIDs {\it Physica C} {\bf 371} 156 -- 172
\bibitem{G03} Greenberg Ya S 2003 Theory of the voltage–current characteristics of high TC asymmetric DC SQUIDs {\it Physica C} {\bf 383} 354 -- 364
\bibitem{GNSM} Greenberg Ya S, Novikov I L, Schultze V and Meyer H-G 2005 The influence of the second harmonic in the current-phase relation on the voltage-current characteristic of high Tc DC SQUID {\it Eur. Phys. J. B} {\bf 44} 57 -- 62

\bibitem{GJTC} Groenbech-Jensen N, Thompson D B, Cirillo M, Cosmelli C 2003 Thermal escape from zero-voltage states in hysteretic superconducting interferometers {\it Phys. Rev. B} {\bf 67} 224505
\bibitem{RD} Romeo F, De Luca R 2004 Effective non-sinusoidal current-phase dependence in conventional d.c. SQUIDs {\it Phys. Lett. A} {\bf 328} 330 -- 334
\bibitem{TDL} Torre G, de Luca R 2013 Persistent currents and magnetic susceptibility of two-junction quantum interferometers {\it Results in Physics} {\bf 3} 179 -- 181
\bibitem{DLFG} De Luca R, Fedullo A and Gasanenko V A 2007 Pertubation analysis of the dynamical behavior of two-junction interferometers {\it Eur. Phys. J. B} {\bf 58} 461 -- 467

\bibitem{PM} Peterson R L and McDonald D G 1983 Voltage and current expressions for a two-junction superconducting interferometer {\it Journ. Appl. Phys.} {\bf 54} 992 -- 996

\bibitem{JLSM} Jaklevic R C, Lambe J, Silver A H and Mercereau J E 1964 Quantum interference effects in Josephson tunneling {\it Phys. Rev. Lett.} {\bf 12} 159

\bibitem{RSJ} Stewart W C 1968 Current-voltage characteristics of Josephson junctions {\it Appl. Phys. Lett.} {\bf 12} 277

\bibitem{MRep} Mukhanov O A 2015 Recent Progress in Digital Superconducting Electronics {\it International Superconductive Electronics Conference (ISEC), 2015 IEEE 15th ISEC, July 6 - 9}

\end{thebibliography}
\end{document}